\documentclass[aps,pre,twocolumn,superscriptaddress,showpacs,floatfix]{revtex4}
\usepackage{dcolumn}
\usepackage{amsmath}
\usepackage{amssymb}
\usepackage{amsthm}
\usepackage{graphicx}
\usepackage{bm}
\usepackage[T1]{fontenc}
\usepackage{color}
\usepackage{url}
\usepackage[bf]{subfigure}
\usepackage{rotating}
\usepackage{scalefnt}
\usepackage{multirow}

\usepackage{scalefnt}

\theoremstyle{definition}
\newtheorem{definition}{Definition}[section]
 
\newcommand{\M}{\mathbf{M}}
\newcommand{\U}{\mathbf{U}}
\newcommand{\X}{\mathbf{X}}
\newcommand{\Xa}{\mathbf{X_1}}
\newcommand{\Xb}{\mathbf{X_2}}
\newcommand{\Y}{\mathbf{Y}}
\newcommand{\J}{\mathbf{J}}
\newcommand{\Om}{\mathbf{0}}
\newcommand{\A}{\mathbf{A}}
\newcommand{\B}{\mathbf{B}}
\newcommand{\C}{\mathbf{C}}
\newcommand{\D}{\mathbf{D}}
\newcommand{\Q}{\mathbf{Q}}
\newcommand{\Sm}{\mathbf{S}}
\newcommand{\Lap}{\mathbf{L}}
\newcommand{\Pt}{\mathbf{P}}
\newcommand{\Po}{\mathbf{R}}
\newcommand{\La}{\mathbf{L_a}}
\newcommand{\Lb}{\mathbf{L_b}}
\newcommand{\Aa}{\mathbf{A_a}}
\newcommand{\Ab}{\mathbf{A_b}}

\DeclareMathOperator{\I}{\mathbf{I}}

\newcommand{\vertiii}[1]{{\left\vert\kern-0.25ex\left\vert\kern-0.25ex\left\vert #1 
    \right\vert\kern-0.25ex\right\vert\kern-0.25ex\right\vert}}
    
\newcommand{\E}[1]{\left \langle #1 \right \rangle}

\begin{document}

\title{A polynomial eigenvalue approach for multiplex networks}

\author{Guilherme Ferraz de Arruda}
\affiliation{ISI Foundation, Via Chisola 5, 10126 Torino, Italy}
\affiliation{Departamento de Matem\'{a}tica Aplicada e Estat\'{i}stica, Instituto de Ci\^{e}ncias Matem\'{a}ticas e de Computa\c{c}\~{a}o, Universidade de S\~{a}o Paulo - Campus de S\~{a}o Carlos, Caixa Postal 668, 13560-970 S\~{a}o Carlos, SP, Brazil.}

\author{Emanuele Cozzo}
\affiliation{Institute for Biocomputation and Physics of Complex Systems (BIFI) \& Department of Theoretical Physics, University of Zaragoza, 50018 Zaragoza, Spain}

\author{Francisco A. Rodrigues}
\affiliation{Departamento de Matem\'{a}tica Aplicada e Estat\'{i}stica, Instituto de Ci\^{e}ncias Matem\'{a}ticas e de Computa\c{c}\~{a}o, Universidade de S\~{a}o Paulo - Campus de S\~{a}o Carlos, Caixa Postal 668, 13560-970 S\~{a}o Carlos, SP, Brazil.}

\author{Yamir Moreno}
\affiliation{Institute for Biocomputation and Physics of Complex Systems (BIFI) \& Department of Theoretical Physics, University of Zaragoza, 50018 Zaragoza, Spain}
\affiliation{ISI Foundation, Via Chisola 5, 10126 Torino, Italy}

\begin{abstract}
We explore the block nature of the matrix representation of multiplex networks, introducing a new formalism to deal with its spectral properties as a function of the inter-layer coupling parameter. This approach allows us to derive interesting results based on an interpretation of the traditional eigenvalue problem. More specifically, we reduce the dimensionality of our matrices but increase the power of the characteristic polynomial, i.e, a polynomial eigenvalue problem. Such an approach may sound counterintuitive at first glance, but it allows us to relate the quadratic problem for a 2-Layer multiplex system with the spectra of the aggregated network and to derive bounds for the spectra, among many other interesting analytical insights.
Furthermore, it also permits to directly obtain analytical and numerical insights on the eigenvalue behavior as a function of the coupling between layers. Our study includes the supra-adjacency, supra-Laplacian and the probability transition matrices, which enable us to put our results under the perspective of structural phases in multiplex networks. We believe that this formalism and the results reported will make it possible to derive new results for multiplex networks in the future. 
\end{abstract}

\maketitle

\section{introduction}

Complex network theory has become one of the main tools for the analysis of complex systems, allowing the representation of a wide range of systems composed by interacting discrete elements \cite{Boccaletti06:PR}. However, real-life systems are also organized in layers, which represent different channels of interaction. In order to incorporate these characteristics, one should work with multilayer networks, which allows for a proper representation of multiplex and interconnected systems \cite{Kivela2014, BoccalettiPR2014, DeDomenico2013}. The introduction of this extra level of complexity also imposes new challenges on the analysis of its structural and dynamical properties. Furthermore, a key element on the analysis of networks is their spectral properties \cite{Mieghem:2011}. In fact, they play an important role in explaining the connection between structure and dynamics. For instance, in epidemic spreading the critical point below which the infection prevalence is null is predicted to be the inverse of the leading eigenvalue of the adjacency matrix, in both, single \cite{Mieghem09} and multiplex networks \cite{Arruda2017}. Additionally, its nature also seems to be connected to those properties \cite{Goltsev2012, Boguna2013, Arruda2017}. Although the literature about the spectra of single-layer networks is well developed \cite{Mieghem:2011}, the theory of spectral properties of multiplex networks is still in its infancy. This motivates us to propose a different formalism aimed at filling this gap.

In this paper, we will consider the matrix representation of multiplex networks, constraining ourselves to finite matrices. First of all, we are interested in weighting differently inter and intra-layer edges. This implies that those matrices will also be a function of the inter-layer coupling parameter, here called $p$. Consequently, the associated eigensystem will be a function of that same parameter. Additionally, the matrix approach is especially interesting in this context since it allows us to directly use linear algebra and spectral graph results already available.

When varying the coupling parameter, a multiplex system might present different structural phases, which are characterized in terms of eigenvalue crossings and eigengaps and are intuitively defined as: (i) decoupled phase, for small values of $p$, where the layers are virtually decoupled and act by themselves, with a negligible interaction between layers, (ii) multiplex/multilayer phase, where the system is coupled and the intra-layer edges play an important role and (iii) the aggregate network phase, where the system behaves as the superposition of all layers. It is clear that a good understanding of the eigenvalues' behavior might be useful since we could move our system into different structural regimes, aiming at different goals such as improved robustness, better performance regarding diffusion or spreading, among many other possible applications.

The different structural phases above are related through the interlacing properties of quotient graphs \cite{Sanchez-Garcia2014, Cozzo2015}. More specifically, in \cite{Sanchez-Garcia2014} the authors showed that the spectra of different scales of a multiplex network (aggregated network, the network of layers and individual layers) characterize the three phases. In practical terms, the interlacing provides us bounds for the spectra \cite{Sanchez-Garcia2014, Cozzo2015}, but also emphasizes that the different scales are intrinsically connected. Indeed, it is impossible to tune the leading eigenvalue of the network of layers without also increasing the leading eigenvalue of the whole multiplex. Furthermore, in \cite{Cozzo2016} the authors characterized multiple topological scales using the supra-Laplacian matrix. More specifically, they analyzed eigengaps to characterize them.

Following a different approach, in \cite{Radicchi2014}, the author evaluated the normalized Laplacian matrix, which, in fact, shares the same set of eigenvalues of the probability transition matrix (see Section \ref{sec:app_P}), and proposed a similar classification. However, it is worth mentioning that in \cite{Radicchi2014}, a different nomenclature was used and a fourth phase was defined. Namely, the proposed structural regimes were: (i) bipartite phase, (ii) decoupled phase, (iii) indistinguishable, where the author argues that the system is topologically and dynamically indistinguishable\cite{Radicchi2014}, and (iv) a mixed phase, (called BD in \cite{Radicchi2014} -- bipartite and decoupled phases) where the layers are structurally and dynamically distinguishable. Although here we do not make a distinction between the regimes (iii) and (iv) and consider both as multiplex regimes, we acknowledge the differences pointed out in \cite{Radicchi2014}. It is also noteworthy that \cite{Radicchi2014} considered structural correlations for the analysis, which is a key ingredient for the reported results. On the other hand, here we focus on uncorrelated networks and a multiplex structure.

The paper is organized as follows. In Section \ref{sec:pep}, we present the polynomial eigenvalue formalism, giving its general definitions and properties in Section \ref{sec:Definition}. Next, we formalize the 2-Layer problem into a quadratic eigenvalue problem in Section \ref{sec:mux_block}, analytically exploring its behavior as a function of the coupling parameter. In Section \ref{sec:spec_p}, obtaining some bounds, in Section \ref{sec:bounds}, and discussing the simplified symmetric problem in Section \ref{sec:hqep}. We present our main applications in Section \ref{sec:app}, where we explore the supra-Laplacian matrix, in Section \ref{sec:app_L}, the supra-adjacency matrix in Section \ref{sec:app_A} and the probability transition matrix in Section \ref{sec:app_P}. To round off this paper, we discuss the physical consequences of our findings, summarize our main results and perspectives in Section \ref{sec:discussion}.

\section{Polynomial eigenvalue problem} \label{sec:pep}

In this section, we formally define the polynomial eigenvalue problem and present some of its fundamental properties. The aim of this section is to generically define our main mathematical object, establishing its basic properties. Thus, this will allow us to properly study the matrices associated with 2-Layer multiplex networks, which will naturally appear as a consequence of a simple manipulation of a linear system describing the network. From this simple approach, we expect to provide a different perspective on the spectral properties of multiplex networks.

\subsection{General definition and properties} \label{sec:Definition}

A matrix polynomial of order $l$ is a matrix-valued function of a complex variable of the form \cite{gohberg1982matrix}
\begin{equation}
 \Po (\lambda) =\sum_{i=0}^{l} \M_i \lambda^i,
\end{equation}
where $\M_0, \M_1,\dots, \M_l$ are $n\times n$ matrices and they are said coefficient matrices. If $\M_l= \I$ the identity matrix the polynomial $\Po$ is said to be monic. The eigenvalues of $\Po$ are the solution to the characteristic equation
\begin{equation}
 \det \left( \Po(\lambda) \right) = 0.
\label{polynomialeigenvalues}
\end{equation}
Right and left eigenvectors are defined by
\begin{eqnarray}
 \Po(\lambda)x = 0, \\
 y^T\Po(\lambda) = 0, 
\end{eqnarray}
where $x$ and $y$ are the right and left eigenvectors associated to the eigenvalue $\lambda$. It reduces to the standard eigenvalue problem when $\Po(l)$ is a monic matrix polynomial with $l=1$ and $\M_0=-\A$, for any matrix $\A$. A generalization of the Jordan form theory to a general matrix polynomial is possible and is briefly presented in the Appendix \ref{sec:app_jordan}.

A particular case of general interest is the Quadratic Eigenvalue Problem (QEP), which is directly related to the 2-layer case of our interest in this work. A quadratic matrix polynomial can be written as \cite{Lancaster:book, gohberg1982matrix, tisseur2001quadratic}
\begin{equation} \label{eq:qep}
 \Q(\lambda) = \A \lambda^2 + \B \lambda + \C.
\end{equation}

Besides, it is worth mentioning that, without loss of generality, we assume in the following that the eigenvectors are unitary. Note, however, that if the right eigenvector $x$ is normalized, then the left eigenvector $y$ is not. In the following, we assume that $x$ is unitary to simplify the equations.

Furthermore, a special class of problems is obtained if $\A$, $\B$ and $\C$ are Hermitian, called Hyperbolic Quadratic Eigenvalue Problem (HQEP) \cite{tisseur2001quadratic}. Unfortunately, most of the problems in our context does not fall in this class.

\section{A general 2-Layer case: a block matricial problem} \label{sec:mux_block}

The general form of any matrix associated to a multiplex network composed by two layers (supra-adjacency, supra-Laplacian and transition matrices for example) can be written as a block matrix. The resulting eigenvalue problem is the following
\begin{equation} \label{eq:block}
 \begin{bmatrix}
  \M_{11} & \M_{12} \\
  \M_{21} & \M_{22} 
 \end{bmatrix}
  \begin{bmatrix}
  v_1 \\
  v_2
 \end{bmatrix}
 = \lambda
 \begin{bmatrix}
  v_1 \\
  v_2
 \end{bmatrix},
\end{equation}
where $\M_{12} = \M_{21}^T$. Interpreting it as a system of equations and isolating $v_1$ on the second row, we have
\begin{equation} \label{eq:v1}
 v_1 = - \M_{21}^{-1} (\M_{22} - \lambda \I)v_2.
\end{equation}
Finally, inserting it in the first row we have
\begin{equation}
 \left[(\M_{11} - \lambda \I)\M_{21}^{-1} (\M_{22} - \lambda \I) - \M_{12} \right] v_2 = 0.
\end{equation}
This expression defines a QEP, whose coefficient matrices are
\begin{eqnarray} \label{eq:mat_coef}
 \A &=& \M_{21}^{-1}, \\
 \B &=& - \left( \M_{11}\M_{21}^{-1} + \M_{21}^{-1} \M_{22} \right), \\
 \C &=& \M_{11} \M_{21}^{-1} \M_{22}  - \M_{12},
\end{eqnarray}
which poses a restriction on the inter-layer coupling matrix $\M_{12}$, i.e., it must be invertible. Additionally, in our context, exchanging $\M_{11}$ and $\M_{22}$ does not change the system, neither the solutions. Note that this operation is equivalent to relabeling the layers. However, if the polynomial of the first is $\Q(\lambda)$, then, for the second it is $\Q(\lambda)^T$. In this way we found a relation between the right and left eigenvectors and these two possible configurations of our system. Formally, such observation implies $x = v_2$ and $y = v_1$. As usual, we consider coupling matrices that are functions of a coupling parameter, $p$, i.e., $\M_{ij}=\M_{ij}(p)$, for $i \neq j$. In fact, throughout this paper we explore how the spectral properties of our network evolve as we change such coupling parameter. As a constraint, we should mention that we only consider finite matrices.

Furthermore, note that $\B$ in equation \ref{eq:mat_coef} is intimately related to the aggregated and the loop-less aggregated networks of the original multiplex network (for more, see \cite{Sanchez-Garcia2014} or Section 2.3.2 of \cite{cozzo2018multiplex}). More specifically, if the coupling matrices $\M_{12}$ and $\M_{21}$ are the identity matrix (or proportional to this matrix), thus $\B = - \left( \M_{11} + \M_{22} \right)$, which is proportional to the loop-less aggregated network. Besides, note that a network of layers in the two-layer multiplex is a simple line graph with two nodes.

\subsection{Spectral analysis as a function of $p$} \label{sec:spec_p}

So far, we have defined our main mathematical objects, making as less constraints as possible. Now we restrict ourselves to diagonal coupling matrices - i.e., multiplex networks - and assume a linear function of the parameter $p > 0$ \footnote{Note that it is necessary to restrict our coupling parameter to $p>0$ since $\A = \M_{12}^{-1}$ and the problem would not be well defined otherwise. Negative coupling parameters would also be possible, but they make less physical sense.}, $\M_{12} = p \D$, where $\D$ is a diagonal invertible matrix (such constraint will be relaxed later). Then, defining the scalar equation that describes each eigenvalue as the product of $\Q(\lambda)$, by its left and right eigenvectors we have
\begin{equation} \label{eq:scalar}
 y^T \Q(\lambda) x = a(y^T,x) \lambda^2 + b(y^T,x) \lambda + c(y^T,x) = 0,
\end{equation}
where $a(y^T,x) = y^T \A x$, $b(y^T,x) = y^T \B x$ and $c(y^T,x) = y^T \C x$. The solution of this equation is given by
\begin{equation} \label{eq:solution}
 \lambda^\pm(x) = \frac{-b(y^T, x) \pm \sqrt{\Delta(y^T, x)}}{2 a(y^T, x)},
\end{equation}
where $\Delta(y^T, x) = b(y^T, x)^2 - 4 a(y^T, x)c(y^T, x)$. Note that for each pair of right and left eigenvectors we have two possible solutions, but just one of them is an eigenvalue of $\Q(\lambda)$. Additionally, differentiating equation~\ref{eq:scalar} by $p$ we obtain information on how the eigenvalues change as $p$ changes. Formally we have
\begin{equation} \label{eq:diff_1}
  \dfrac{\partial y^T \Q (\lambda) x}{\partial p} = y^T \dfrac{\partial \Q (\lambda)}{\partial p} x + y^T \Q(\lambda) \dfrac{dx}{dp} + \dfrac{dy}{dp} \Q(\lambda) x = 0,
\end{equation}
where
\begin{equation} \label{eq:diff_2}
 \dfrac{\partial \Q (\lambda)}{\partial p} = 2 \lambda \A \dfrac{d \lambda}{dp} + \dfrac{d \lambda}{dp} \B + \lambda \dfrac{\partial \B}{\partial p} + \dfrac{\partial \C}{\partial p}.
\end{equation}
Note that the eigenvalues and eigenvectors are also a function of $p$. For continuity, two different eigenvalues may cross each other when varying $p$. Observe that for non-crossing points the relations $\dfrac{dy^T}{dp} \Q(\lambda) x = 0$ and $y^T \Q(\lambda) \dfrac{dx}{dp} = 0$ holds, since the derivatives are bounded for non-crossing points. However, on the crossings we have two eigenvectors associated to the same eigenvalue, which imply two solutions for the derivatives. Then, isolating the derivative of $\lambda$ we have
\begin{equation} \label{eq:dldp}
 \dfrac{d \lambda}{dp} = \frac{y^T \left( -\lambda \dfrac{\partial \B}{\partial p} - \dfrac{\partial \C}{\partial p}\right) x}{y^T \left( 2 \lambda \A + \B \right) x}.
\end{equation}
Such relation can be applied to drive a system through different regimes. For instance, considering the adjacency matrix, one can use this equation in order to chose an edge or set of edges to be removed (or weighted) in order to optimally reduce or increase the leading eigenvalue and consequently the critical point of spreading processes, such as epidemic spreading. Obviously the matrix under study depends on the process. Another application is to design a numerical method to follow the correct eigenvalues as a function of $p$ in a problem that might present eigenvalues crossings \cite{Mieghem2015, Arruda2017}.

\subsection{Bounds} \label{sec:bounds}

Aiming to find bounds to equation~\ref{eq:solution} we study the scalar polynomial defined by $x^T \Q(\lambda) x = 0$, where $x$ is an eigenvector (left or right), which guarantees that the polynomial is equal to zero. In order to simplify the problem we multiply $\Q(\lambda)$ by $\D$, obtaining a monic matrix polynomial , then we must bound the terms $b(x^T,x)$ and $\Delta(x^T, x)$, which allow us to bound both solutions. Those terms can be bounded by the numerical range of the matrices to which they are related. The numerical range is formally defined for any matrix $\X$ as $F(\X) = \{ x^T\X x : x \in \mathbb{C} \text{ and } x^Tx = 1 \}$. Additionally, $\sigma(\X) \subseteq F(\X)$, where $\sigma(\X)$ is in the set of eigenvalues of $\X$. Moreover, if $\X$ is an Hermitian matrix $x^T\X x$ is the Rayleigh quotient of $\X$, which implies $\lambda_1(\X) \leq x^T\X x \leq \lambda_N(\X)$. Finally, to bound a non-Hermitian matrix we use the relation of the spectral norm and the numerical range, given as $\frac{1}{2} \vertiii{\X}_2 \leq r(\X) \leq \vertiii{\X}_2$, where $r(\X)$ is its numerical radius, defined as $r(\X) =  \underset{\left\Vert x \right\Vert_2 = 1}{\max} |x^* \X x| = \max \{ |z| : z \in F(\X)\}$. 

First, consider the term $b(x^T,x)$, which is bounded by
\begin{equation}
 - \vertiii{\B}_2 \leq b(x^T,x) \leq \vertiii{\B}_2,
\end{equation}
however, in many cases, $\B$ is an Hermitian matrix, allowing us to improve this bound to
\begin{equation}
 \lambda_{\min} (\B) \leq b(x^T,x) \leq \lambda_{\max} (\B)
\end{equation}
More precisely, observe that $\B$ is often related to the aggregated network, connecting both scales.

Next, we evaluate $\Delta(x^T,x) = (x^T \B x)^2 - 4 x^T \C x$. Firstly, we analyze the term $(x^T \B x)^2$, by observing that: (a) $\min \{ \mu_i \} \leq x^T\B x \leq \max\{ \mu_i \}$, (b) $\min\{\mu_i^2\} \leq x\B^2 x \leq \max\{\mu_i^2\}$ and (c) $\min\{|\mu_i|\}^2 \leq (x^T\B x)^2 \leq \max\{|\mu_i|\}^2$, since $\min\{\mu_i^2\} = \min\{|\mu_i|\}^2$, hence, from (b) and (c), bounding $(x^T\B x)^2$ is equivalent to bound $x\B^2 x$. Secondly, we can factorize $\Delta(x^T,x) = x^T (\B^2 - 4 \C) x$ and defining the matrix $\Delta = \B^2 - 4 \C$, we can focus on the problem $x^T \Delta x$ instead of the initial definition of $\Delta(x^T,x)$, since both have the same bounds. Besides, since in most of the problems on networks we are dealing with symmetric matrices (undirected networks), we might also impose that  $\Delta(x^T,x) \geq 0$ because we already know that the spectra is real in this case. consequently, we have 
\begin{equation}
 0 \leq \Delta(x^T,x) \leq \vertiii{\Delta}_2.
\end{equation}
Observe that those bounds can be further improved when applied to the analysis of particular matrices (supra-adjacency, supra-Laplacian, and probability transition) since their particularities also impose constraints on the solutions and could be explored to improve the bounds.

\subsection{Comments on symmetric problems: HQEP} \label{sec:hqep}

As previously mentioned, if $\A$, $\B$ and $\C$ are Hermitian, we have a special class of problems called Hyperbolic Quadratic Eigenvalue Problem (HQEP) \cite{tisseur2001quadratic}. The HQEP has interesting properties, for instance, if $x$ is a right eigenvector associated with the eigenvalue $\lambda$, then it is also a left eigenvector of the same eigenvalue\cite{tisseur2001quadratic}. In order to take advantage of those properties one can interpret the original problem as an HQEP plus asymmetric perturbation. Thus, the matrix polynomial defined by the matrix coefficients in Equation~\ref{eq:mat_coef} is not symmetric in most cases. However, a class of problems that arise naturally is defined by $\M_{12} = p\I$ and in this case the matrices $\A$ and $\B$ are Hermitian. Observe that $\C$ still might be asymmetric. However, we can use the Toeplitz decomposition \cite{Horn:2012:MA} in order to analyze a simplified problem. Such decomposition states that any square matrix can be uniquely written as the sum of an Hermitian ($\X = \X^*$) and a skew Hermitian matrix ($\X = - \X^*$) as $\X = \frac{1}{2} (\Xa + \Xb^*) + \frac{1}{2} (\Xa - \Xb^*)$. This allows us to decompose $p\C = \frac{1}{2} (\M_{11} \M_{22} + \M_{22} \M_{11}) + \frac{1}{2} (\M_{11} \M_{22} - \M_{22} \M_{11}) + p^2\I$. In this way we can re-write our QEP into two parts, one composed by Hermitian matrices, which is a HQEP, and a skew Hermitian matrix, that can be interpreted as a perturbation. The natural consequence from the perturbation theory is that the matrix $p\C$ of the HQEP is perturbed by $\frac{1}{2} (\M_{11} \M_{22} - \M_{22} \M_{11})$ and such matrix norm goes to zero as the layers are more similar. From the Bauer and Fike theorem \cite{Horn:2012:MA}  we can write a quality function for the approximation of the perturbed matrix $\C$ as
\begin{equation}
 \left| \lambda - \hat{\lambda} \right| \leq \kappa(\U) \vertiii{\frac{1}{2} (\M_{11} \M_{22} - \M_{22} \M_{11})},
\end{equation}
where $\hat{\lambda}$ is the eigenvalue of $\C = \mathbf{C_{H} + \mathbf{C_{S}}}$, $\mathbf{C_H} = \U \Lambda \U^{-1}$ and $\kappa(\cdot)$ is the condition number with respect to the matrix norm $\vertiii{\cdot}$. Considering the spectral norm $\vertiii{\cdot}_2$ we have $\kappa(\X) = \left| \frac{\sigma_{\max}(\X)}{\sigma_{\min}(\X)} \right|$. If $\kappa(\U)$ is near 1, small perturbations imply small changes on the eigenvalues. On the other hand, large values of  $\kappa(\U)$ suggest a poor approximation. Observe that such analysis concerns only the matrix $\C$ and not the whole QEP, however, it can be an estimate of the quality of the approximation and show that the general solution interpolates between a HQEP and a general QEP.

In addition to the HQEP properties, the perturbation analysis also emphasizes an important multiplex property. We must note that the more similar the layers are, the closer to zero the norm $\vertiii{\frac{1}{2} (\M_{11} \M_{22} - \M_{22} \M_{11})}$ is. Moreover, we also have another criteria which is based on the commutativity of the matrices $\M_{11}$ and $\M_{22}$. Moreover, observe the role of correlations in this approximation. If both layers are identical, they are obviously correlated and the problem is symmetric.

\subsection{Limits for sparse inter-layer coupling: singular $\D = \M_{12} = \M_{21}$} \label{sec:sparse}

So far we have assumed a node-aligned multiplex, i.e., a multiplex network in which each node has a counterpart on every layer\cite{Kivela2014}, fulfilling the invertibility of $\M_{12}$, which is necessary to formally define the problem, however, we can use the limit of ${\D_{ii} = \epsilon \rightarrow 0}$ to obtain an approximation of the sparse coupling. Observe that equation~\ref{eq:qep} can be analyzed in two different steps, first calculating the limit of decoupled edges and secondly the rest of the system. The first limit is analyzed as follows. From ~\ref{eq:qep} the absent edges are factorized as
\begin{equation} \label{eq:scalar_sparse}
\begin{split}
  &p^{-1} \epsilon^{-1} \tilde{\D}(i) \lambda^2 - p^{-1} \epsilon^{-1} \left( \M_{11}\tilde{\D}(i) + \tilde{\D}(i)\M_{22} \right) \lambda + \\ &p^{-1} \epsilon^{-1} \M_{11} \tilde{\D}(i) \M_{22}  - p \D  = 0,
\end{split}
\end{equation}
where $\tilde{\D}(i) = \epsilon\D^{-1}$. Multiplying equation~\ref{eq:scalar_sparse} by $p \epsilon$ and using the following limit
\begin{equation}
 \lim_{\epsilon \rightarrow 0} \left[\epsilon \D^{-1} \right]_{jj} = \begin{cases}
                                    1  \hspace{0.5cm} \text{if} \hspace{0.5cm} \left[\D^{-1} \right]_{jj} \in O\left( \epsilon \right), \\
                                    0  \hspace{0.5cm} \text{otherwise}.
                                   \end{cases}
\end{equation}
we have 
\begin{equation} \label{eq:qep_sparse}
\begin{split}
  \tilde{\D}\lambda^2 -\left( \M_{11}\tilde{\D} + \tilde{\D}\M_{22} \right)  \lambda + \M_{11} \tilde{\D} \M_{22} = 0,
\end{split}
\end{equation}
where the term of order $p \epsilon \D$ vanishes in the limit of $\epsilon \rightarrow 0$. Observe that $\tilde{\D} = \lim_{\epsilon \rightarrow 0} \left[\epsilon \D^{-1} \right] = \I$ if both layers are decoupled and the polynomial equation can be factorized as $\left( \M_{11} - \lambda \I\right)\left( \M_{22} - \lambda \I\right) = 0$, whose solutions are the union of the solution of the standard eigenvalue problem of each layer. An important observation is that the number of nodes that are not connected to the other layer is also the number of eigenvalues that do not change as a function of $p$.

Equation~\ref{eq:qep_sparse} presents the solution for nodes that do not have any counterpart on the other layer. In order to calculate the remaining solutions we have to redefine the original problem in terms of the Moore -- Penrose pseudoinverse, denoted by $\X^{\dagger}$, for a matrix $\X$. Denoting by $\bar{\D} = p^{-1} \D^{\dagger}$ we have $\bar{\D}_{jj} = p^{-1} \D_{jj}^{-1}$ if $\D_{jj} \neq 0$ and $\bar{\D}_{jj} = 0$ otherwise. Note that the zeros of $\bar{\D}_{jj}$ are ones in $\tilde{\D}_{jj}$. For the sake of simplicity, in the following we assume that $\M_{12}$ is invertible, however, the strategy mentioned above can be applied if it is not the case. From the computational point of view, we can reduce the cost to calculate the whole spectra as a function of a closed range of $p$ by separating it into two components, where a subset is constant and the remaining subset varies.

\section{Applications} \label{sec:app}

We next apply our main formalism to study the supra-Laplacian and the supra-adjacency matrices. For the sake of completeness, let us explicitly define the supra-adjacency matrix in terms of its block matrices (adjacency matrix of the individual layers). Formally, the supra-adjacency matrix is defined as
\begin{equation}
 \A = 
  \begin{bmatrix}
  \Aa & p\I \\
  p\I & \Ab 
 \end{bmatrix},
\end{equation}
where we weight differently the intra and inter-layer edges. The definition of the supra-Laplacian matrix is
\begin{equation}
 \Lap = \hat{\D} - \A = 
  \begin{bmatrix}
  p\I + \La & -p\I \\
  -p\I & p\I + \Lb 
 \end{bmatrix},
\end{equation}
where $\hat{\D}$ is a diagonal matrix whose elements are $\D_{ii} = \sum \A_{ij}$ and the Laplacian matrices of the individual layers are denoted as $\La$ and $\Lb$. Those matrices are related to many dynamical processes. The supra-Laplacian is used to describe diffusion and synchronization of coupled oscillators, while the supra-adjacency matrix is intimately related to epidemic and information spreading. It is also noteworthy that many structural metrics are also directly extracted from the spectral properties of those matrices. For instance, the communicability, which can be easily written as a matrix function, or more specifically, as the exponential of the adjacency or supra-adjacency matrix. Here we are going to focus on the spectral properties of these matrices, and their behavior as a function of the coupling parameter $p$ under different conditions. 
 
In addition to the supra-adjacency and supra-Laplacian matrix, we also analyze the probability transition matrices in Section \ref{sec:app_P}, which can be used to describe classical random walks on networks. The analysis of such a matrix is left to the last section since it is mainly numerical. Note that the probability transition matrix has a well bounded spectra where $1 = \lambda_1 \geq \lambda_2 \geq \lambda_3 \geq ... \geq \lambda_N \geq -1$ \cite{Zhang2011}. This characteristic imposes an extra challenge on the derivation of the bounds. Although we could not improve those bounds, we report an interesting spectral behavior found numerically.

\subsection{Supra-Laplacian matrix} \label{sec:app_L}

The simplest supra-Laplacian matrix can be built considering a diagonal coupling matrix $M_{12} = -p \I$, where each node has a counterpart on the other layer and the coupling is homogeneous. This implies that the QEP is defined with the following coefficient matrices
\begin{eqnarray} \label{eq:lap_mat_coef}
 \A &=& \I, \\
 \B &=& - \left( \La + \Lb + 2p \I \right), \\
 \C &=& \La \Lb + p\left( \La + \Lb\right).
\end{eqnarray}
It is noteworthy that the aggregated network, $\Lap^{+} = \La + \Lb$, appears naturally under this formalism. This is interesting since it is physically understandable. On the other hand, the term $\La \Lb$, in the definition of $\C$, is of not so direct interpretation. This system presents a structural transition, which can be directly derived from our formalism. This derivation is presented in Section \ref{sec:Lap_st}. Additionally, we can also obtain bounds for the spectra using the ideas discussed in Section \ref{sec:bounds}. Those improved bounds are derived in Section \ref{sec:Lap_bounds}, where we use the particular properties of a Laplacian matrix to improve our previous results. In Section \ref{sec:Lap_funct_p}, we evaluate the spectra of the supra-Laplacian matrix as a function of $p$ and also compare our previous results with sparse and heterogeneous couplings. Specifically, on the heterogeneous case we consider a coupling matrix $M_{12} = -p \D$, where $\D$ is a diagonal matrix. The QEP of such matrix is defined by $\A = \D^{-1}$, $\B = - \left( \La\D^{-1} + \D^{-1}\Lb + 2p \I \right)$ and $\C = \La \D^{-1} \Lb + p\left( \La + \Lb\right)$. The analysis of such QEP is not trivial, since the matrices are not symmetric, however we can explore it numerically and compare with the homogeneous case,  $M_{12} = -p \I$.

\subsubsection{Structural transitions} \label{sec:Lap_st}

Firstly, we discuss the structural transition presented in~\cite{Radicchi2013} on the Laplacian matrix. Here we calculate the exact transition points using the QEP formulation. We can easily derive such transition points using our formalism. It is noteworthy that those transition points were also calculated in~\cite{Mieghem2015} using two different methods: eigenvalue sensitivity analysis and a Shur’s complement approach. Both derivations are quite complicated, contrasting with our approach, where the solutions are given using simple arguments. Note, however, that our approach presents a different expression if compared to the method presented in~\cite{Mieghem2015}, but both expressions yield the same final result. We do not prove the equivalence mathematically, but verified their equivalence numerically. 

To begin with, it is well known that $\lambda = 2p$ is an eigenvalue of the supra-Laplacian and the crossing points are a consequence of this eigenvalue crossing the bounded part of the supra-Laplacian spectra, producing the so-called structural transitions. In this way, from our definition of QEP, we have that
\begin{equation} \label{eq:detQ2p}
  \det \left( \Q(2p) \right) = \det \left( \La + \Lb \right) \det \left( \Lb \La \left( \La + \Lb\right)^\dagger - p\I \right),
\end{equation}
which has two possible solutions: (i) $\det \left( \La + \Lb \right) = 0$, which is always true, since the sum of two Laplacian matrices is also the Laplacian of the aggregated network and also has determinant equal to zero and (ii) the solution of $\det \left( \Lb \La \left( \La + \Lb\right)^\dagger - p\I \right)$, which are the crossing points or eigenvalues of multiplicity larger than one. Since it is also an eigenvalue problem in terms of $p$, we have that the crossing points are expressed as $p^* = \lambda_i \left( \Lb \La \left( \La + \Lb\right)^\dagger \right)$. There are $N$ possible values of $p$ that solve Equation \ref{eq:detQ2p}, each one representing one crossing. The first crossing is trivial, at $p=0$, the second is the one called structural transition in \cite{Radicchi2013}, which is relevant for some dynamical processes \cite{GomezPRL2013}. As said before, this expression is different from the previous one presented in the literature, however both give the same result.

\subsubsection{Bounds} \label{sec:Lap_bounds}

\begin{figure}[!t]
\begin{center}
\includegraphics[width=0.98\linewidth]{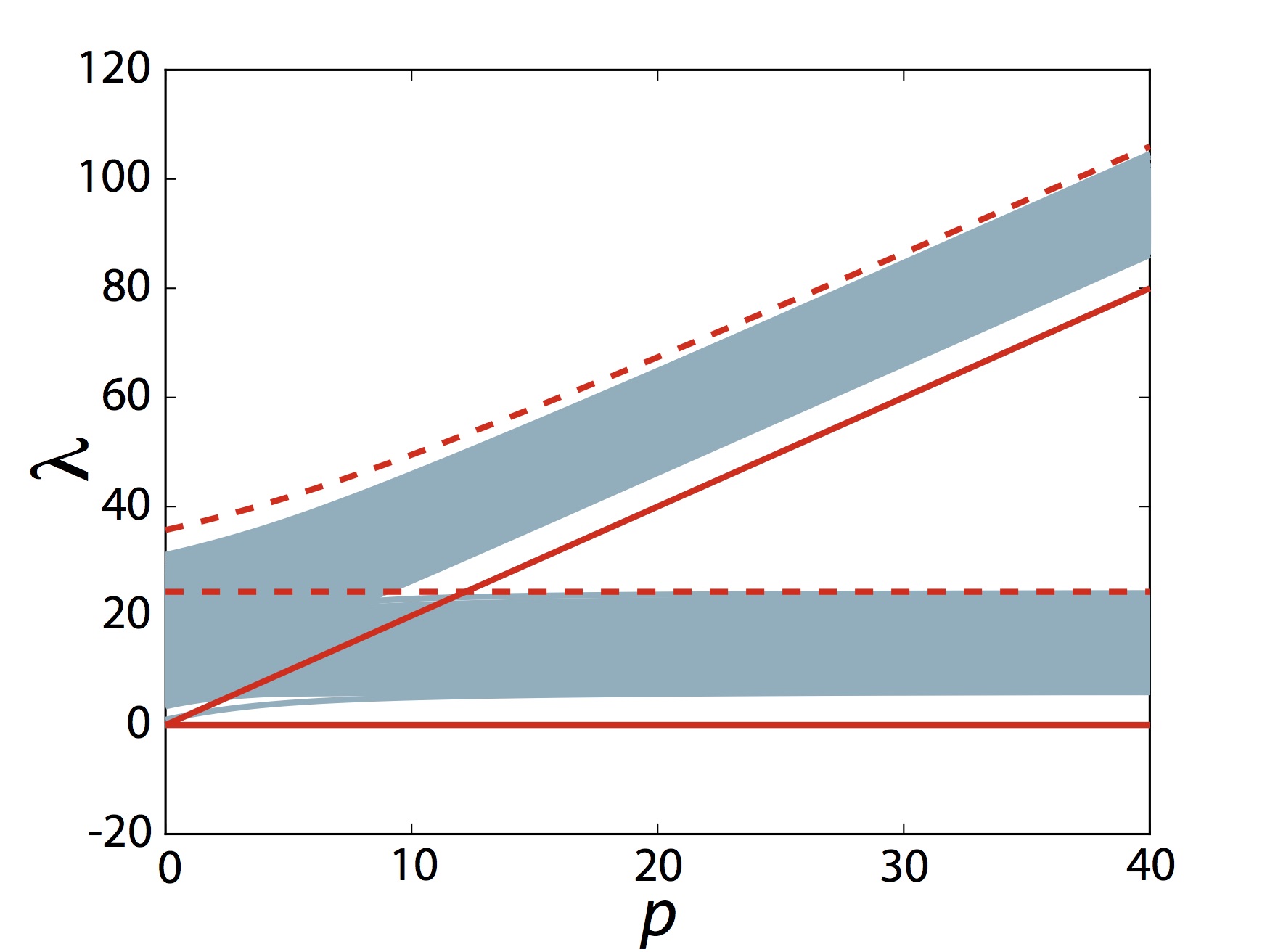}
\end{center}
\caption{Evaluation of the eigenvalues $\lambda(\Lap)$ as a function of the coupling parameter $p$ of a multiplex network composed by two Erd\"os -- Renyi layers with $n = 10^3$ nodes. The first layer has average degree $\E{k} = 12$, while the second has $\E{k} =16$. The continuous lines are the upper bounds, while the dashed lines are the lower bounds.}
\label{fig:BoundsL}
\end{figure}

Here, exploiting the ideas presented in Section \ref{sec:bounds}, we improve the former bounds using specific Laplacian properties, such as its semi-positiveness. Thus, the QEP of the supra Laplacian can be bounded considering the individual bounds of $\B$, which is a semi-positive definite Hermitian matrix, leading to
\begin{equation}
 2p \leq -b(x^T,x) \leq 2p + \lambda_{\max} (\La + \Lb).
\end{equation}
Besides, the discriminant function is also bounded by 
{
\scalefont{0.9}
\begin{eqnarray} \label{eq:bounds_delta_l_0}
  &\min \{ x^T \left( \left( \La - \Lb \right)^2 - 2p \left( \La + \Lb \right) + 4 p^2 \I \right) x \} \leq \Delta(x^T,x) \nonumber \\
  &\Delta(x^T,x) \leq \max \{ x \left( \left( \La - \Lb \right)^2 \right) + 4 p^2 \I \},
\end{eqnarray}}
where the upper bound can be defined as a function of the spectral properties of $\left( \La - \Lb \right)^2$. On the other hand, regarding the lower bound, it can be improved by realizing that the matrix $\Delta = \left( \La - \Lb \right)^2 - 2p \left( \La + \Lb \right) + 4 p^2 \I$, defined on Section \ref{sec:bounds}, is semi-positive definite for undirected networks, $\Delta \succeq 0$. In this way, $\left( \La - \Lb \right)^2 - 2p \left( \La + \Lb \right) + 4 p^2 \I \succeq 0$, hence $\left( \La - \Lb \right)^2 + 4 p^2 \I \succeq 2p \left( \La + \Lb \right)$, implying that $\lambda_i \left( \left( \La - \Lb \right)^2 + 4 p^2 \I \right) \geq \lambda_i \left( 2p \left( \La + \Lb \right)\right)$\footnote{In addition, lets recall that if $\M_1 - \M_2 \succeq 0$, and $\M_1$ and $\M_2$ are semi-positive matrices, with $\M_1 \succeq \M_2$, then $\lambda_i(\M_1) \geq \lambda_i(\M_2)$, where the eigenvalues are in descending order.}. From these properties, we can establish the lower bound as $4p^2$. Formally,
\begin{eqnarray} \label{eq:bounds_delta_l}
  4p^2 \leq \Delta(x^T,x) \leq \lambda_{\max} \left( \left( \La - \Lb \right)^2 \right) + 4 p^2
\end{eqnarray}
The previous bounds imply that in the asymptotic analysis formalism we have $\Delta(x^T,x) \in  \Theta(p^2)$. Moreover, observe that the lower and the upper bounds converge to each other as the layers become similar. On the extreme case of identical layers we have $\Delta(x^T,x) = 4p^2$. Finally, combining the formerly obtained bounds we have,
\begin{eqnarray} \label{eq:bounds_sol1_l}
  & 0  \leq \lambda^-(x^T,x) \leq \frac{1}{2} \lambda_{\max}(\La+\Lb)
\end{eqnarray}
and
\begin{eqnarray} \label{eq:bounds_sol2_l}
  &2p \leq\lambda^+(x^T,x) \leq \nonumber \\
  &\leq p + \frac{\lambda_{\max}(\La+\Lb)}{2}  + \frac{\sqrt{\lambda_{\max}\left(\left(\La - \Lb \right)^2 \right) + 4 p^2}}{2}.
\end{eqnarray}
Interestingly, these bounds can be analyzed in terms of their asymptotic behavior (approximation), where for a sufficiently large value of $p$ they can be approximated to 
\begin{eqnarray}
 0 \leq &\lambda^-(x^T,x)& \leq \frac{\lambda_{\max}(\La+\Lb)}{2}, \\
 2p \leq &\lambda^+(x^T,x)& \leq 2p + \frac{\lambda_{\max}(\La+\Lb)}{2}.
\end{eqnarray}
Observe that from the asymptotic point of view we have $\lambda^-(x) \in \Theta(1)$ and $\lambda^+(x) \in \Theta(p)$. 

As an example, in Figure~\ref{fig:BoundsL} we present the evaluation of the eigenvalues as a function of the coupling parameter $p$ of a multiplex network composed by two Erd\"os Renyi layers with $n = 10^3$ nodes. The first layer has an average degree $\E{k} = 12$, while the second has $\E{k} =16$.

\subsubsection{Spectral properties as a function of the coupling $p$} \label{sec:Lap_funct_p}

\begin{figure}[!t]
\begin{center}
\includegraphics[width=0.98\linewidth]{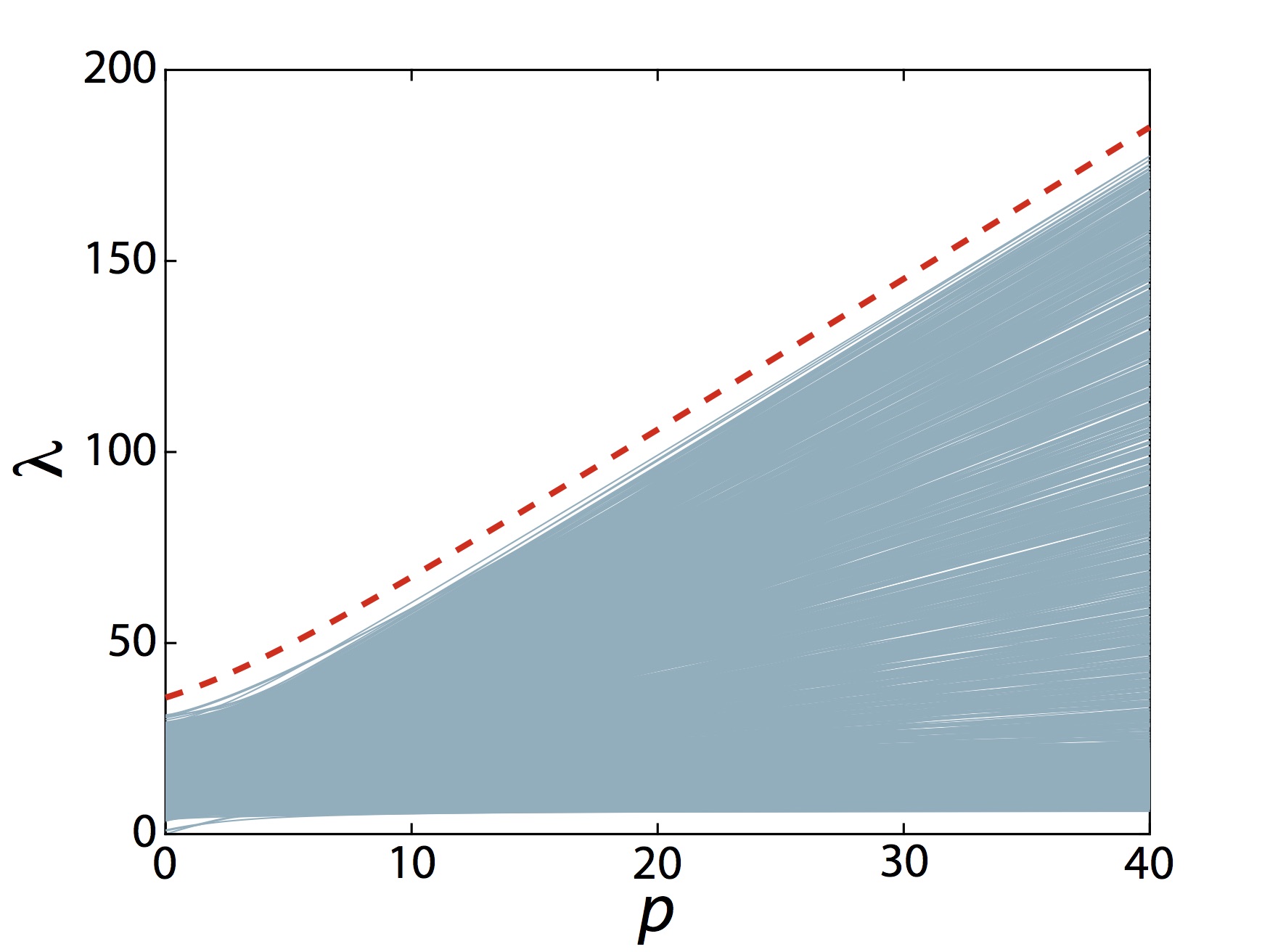}
\end{center}
\caption{Evaluation of the eigenvalues $\lambda(\Lap)$ as a function of the coupling parameter $p$ of a multiplex network composed by two Erd\"os -- Renyi layers with $n = 10^3$ nodes. The first layer has average degree $\E{k} = 12$, while the second has $\E{k} =16$. The coupling matrix is $\D = p \left( \frac{n}{\sum_i^n i}\text{diag}(1,2,...,n) \right)$. The dashed line is the adapted upper bound.}
\label{fig:BoundsLnonidentity}
\end{figure}

In this section, we focus on the behavior of the eigenvalues as a function of the coupling parameter, $\lambda_i(p)$. First of all, we apply the concepts of Section \ref{sec:spec_p} regarding the derivative of $Q(\lambda)$. Consider the simplest case, where $\D = \I$. In such a case, we have a monic polynomial matrix, where $\B$ depends on the aggregated network, which is semi-positive definite. Besides, $\C$ is a matrix that contains the product of both layers and accounts for similarities between them. In this way, Equation \ref{eq:dldp} can be expressed as
\begin{equation} \label{eq:dldp_L}
  \dfrac{d \lambda}{dp} = \frac{\left( 2 \lambda y^Tx - \hat{b}(y^T,x) \right)}{\left( 2 (\lambda - p) y^Tx - \hat{b}(y^T,x)\right)},
\end{equation}
where $\hat{b}(y^T,x) = y^T(\La + \Lb)x$ and $y^Tx = \cos(\theta)$ is the cosine of the angle between left and right eigenvectors of our QEP. Observe that part of the spectra has $\dfrac{d \lambda}{dp} \rightarrow 0$, while the other part has $\dfrac{d \lambda}{dp} \rightarrow 2$ as $p$ increases, which can be proved as follows. Firstly, suppose that $\lambda$ is constant as a function of $p$, then $\dfrac{d \lambda}{dp} \rightarrow 0$ because the denominator grows as a function of $p$ and the numerator is bounded, as supposed. Secondly, suppose that $\lambda$ grows with $p^r$, where $r < 1$. In this case $\dfrac{d \lambda}{dp} \rightarrow 0$, by the same arguments as before, since the linear function of the denominator dominates it. However, if $r = 1$ we have $\dfrac{d \lambda}{dp} \rightarrow 2$, since both, the numerator and the denominator, grow linearly. Finally, with $p^r$, where $r > 1$, both the numerator and denominator are dominated by $p^r$, which imply that for large $p$ the derivative $\dfrac{d \lambda}{dp} \rightarrow 1$, which is also a contradiction, since it was supposed to be a linear function of $p$. In this way, we conclude that the derivatives of $\lambda$, for large values of $p$, cannot grow faster than linearly and their growth will be one of two values, $0$ or $2$. These results are in agreement with the previously obtained bounds. Additionally, as an example, in Figure~\ref{fig:BoundsL} we also observe such a behavior. 

Although for the simplified case we have two possible solutions at large $p$, observe that the above arguments fail for the case of general coupling matrix. From Equation~\ref{eq:dldp} and the definition of the Laplacian QEP we conclude that only the denominator of Equation \ref{eq:dldp} changes for a different choice of $\D$, since the terms that have dependencies on $\D$ vanish in the partial derivatives of the numerator. The denominator follows the general form $y^T \left( 2 \lambda \D^{-1} - 2p \I - \La\D^{-1} - \D^{-1}\Lb \right) x$. In this way, different coupling weights can change the behavior of each eigenvalue differently for large $p$. For instance, if $\D = \I$ the spectral distribution for large $p$ is bimodal, however, if $\D = p \left( \frac{n}{\sum_i^n i}\text{diag}(1,2,...,n) \right)$ \footnote{$\text{diag}(\D)$ has identically spaced numbers and unitary average, due to the term $\frac{n}{\sum_i^n i}$, allowing the comparison with any other figure on this paper.}, this behavior changes completely and the eigenvalues change with different rates, presenting a ``continuous'' bulk. This argument is valid for infinity size networks, since for finite size networks for a large $p$ gaps between eigenvalues may appear due to different rates of growth (as a function of $p$). An example of this is shown in Figure~\ref{fig:BoundsLnonidentity}. Furthermore, we also found an empirical function that seems to bound the spectra as a function of $p$ in this experiment. The lower bound is trivial, since it is a semi-positive definite matrix. The upper bound can be obtained correcting $\tilde{p} = \max\{ \text{diag} (\D)\} p$, hence 
{ \scalefont{0.9}
\begin{equation} 
0 \leq \lambda \leq \tilde{p} + \frac{\lambda_{\max}(\La+\Lb)+ \sqrt{\lambda_{\max}\left(\left(\La - \Lb \right)^2 \right) + 4 \tilde{p}^2}}{2}. \nonumber
\end{equation}}
From Figure~\ref{fig:BoundsLnonidentity} we observe that such bound is not as close to the largest eigenvalue as the homogeneous case.

In addition to a non-homogeneous coupling matrix, the last case studied is the sparse coupling. The analytical part of this study was presented in Section~\ref{sec:sparse}. As predicted, each uncoupled node implies a pair of eigenvalues that does not depend on $p$. Due to the nature of the Laplacian matrix, where just one eigenvalue varies with $p$, while the other remains bounded, the set of bounded eigenvalues increases by one. For example, if we have $\tilde{n}$ uncoupled nodes, the bounded part have $n + \tilde{n}$ eigenvalues, while the ``unbounded'' part have $n-\tilde{n}$ eigenvalues. Note that the upper bound for the bounded part is not $\frac{1}{2} \lambda_{\max}(\La+\Lb)$ anymore. However, the general upper bound for $\D = \I$ seems to be also an upper bound for the sparse problem, as we numerically verified. The figures for these are not shown since they are visually similar to Figure~\ref{fig:BoundsL}.

\subsection{Supra-adjacency matrix} \label{sec:app_A}

Similarly to the supra-Laplacian case, here we also begin with the simplest case, i.e., the diagonal homogeneous coupling, and increase the level of complexity considering heterogeneous inter-layer weights and sparsity. Thus, in the simplest case we have $M_{12} = p \I$, therefore, the QEP, Equation \ref{eq:qep}, is defined by the following coefficient matrices
\begin{eqnarray} \label{eq:adj_mat_coef}
 \A &=& \I, \\
 \B &=& - \left( \Aa + \Ab  \right), \\
 \C &=& \Aa \Ab - p^2 \I.
\end{eqnarray}
Note that, in a way similar to the Laplacian, $\B$ is also defined in terms of the aggregated network. On the other hand, the physical interpretation of $\C$ is still difficult due to the product $\Aa \Ab$. 

In Section \ref{sec:A_bounds} we improve the bounds proposed in Section \ref{sec:bounds}. Then, in Section \ref{sec:A_funct_p}, we evaluate the spectral properties of the supra-adjacency matrix as a function of $p$ in three different contexts: (i) diagonal homogeneous coupling, (ii) diagonal heterogeneous coupling and (iii) sparse diagonal homogeneous coupling. Note that, in order to analyze the heterogeneous coupling we must consider the general QEP with the following coefficient matrices $\A = \D^{-1}$, $\B = - \left( \Aa\D^{-1} + \D^{-1}\Ab  \right)$ and $\C = \Aa \D^{-1} \Ab - p^2 \D$.

\subsubsection{Bounds}  \label{sec:A_bounds}

\begin{figure}[!t]
\begin{center}
\includegraphics[width=0.98\linewidth]{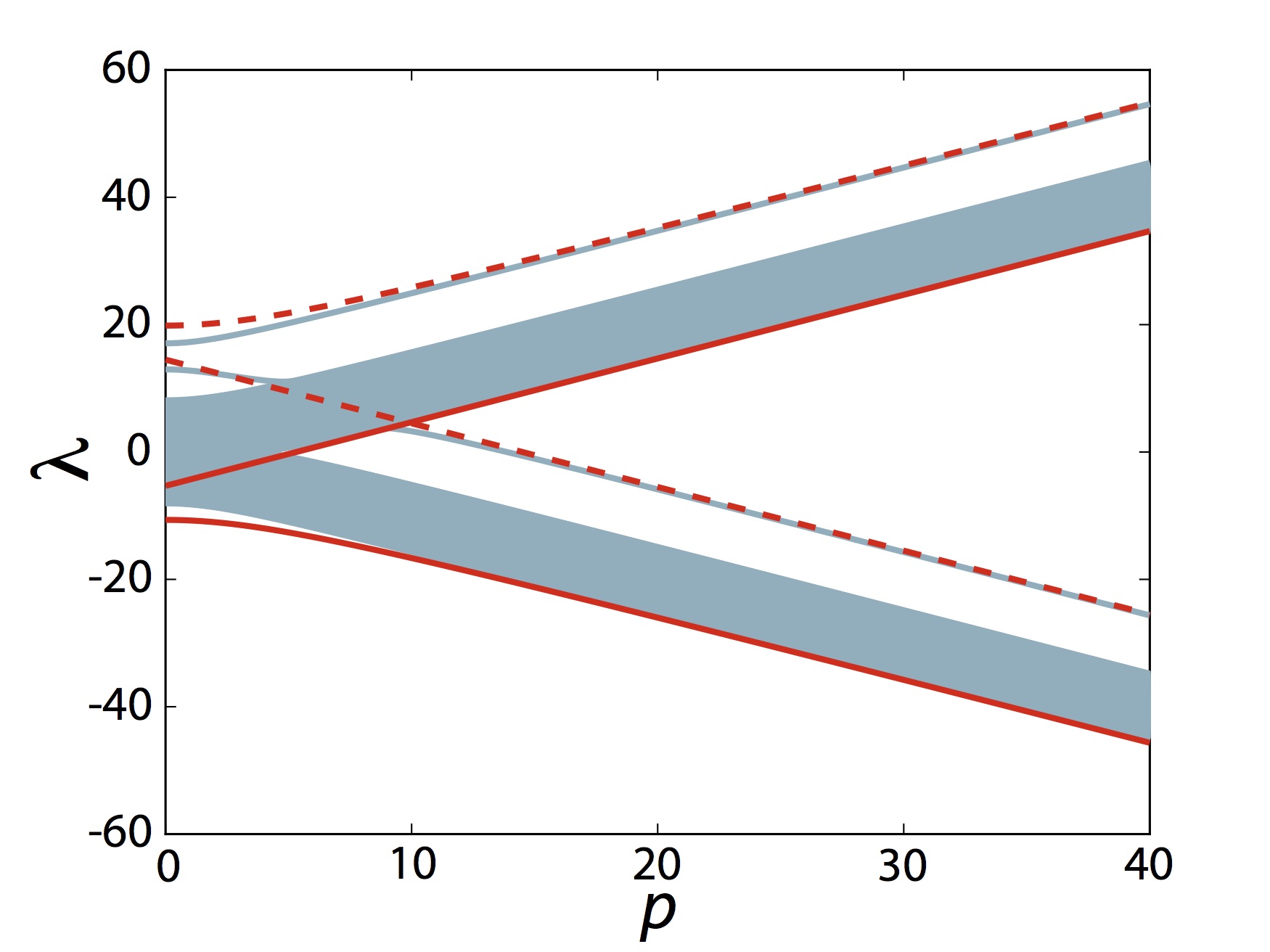}
\end{center}
\caption{Evaluation of the eigenvalues $\lambda(\A)$ as a function of the coupling parameter $p$ of a multiplex network composed by two Erd\"os -- Renyi layers with $n = 10^3$ nodes. The first layer has average degree $\E{k} = 12$, while the second has $\E{k} =16$. The dashed lines are the upper bounds, while the continuous lines are the lower bounds.}
\label{fig:BoundsA}
\end{figure}

Similarly to the analysis performed for the supra-Laplacian, here we also extend the ideas presented in Section
\ref{sec:bounds} to the supra-adjacency matrix. First of all, regarding the diagonal heterogeneous coupling case, $\D = \I$, we can also find bounds for the spectral distribution of the adjacency matrix. Beginning with $\B$, we can bound it based on its eigenvalues as
\begin{equation} \label{eq:bounds_b_A}
 \lambda_{min} (\Aa + \Ab) \leq -b(x) \leq \lambda_{max} (\Aa + \Ab).
\end{equation}
Interestingly, those are the eigenvalues of the aggregated network, which have a clear physical meaning. Similarly, for the discriminant we have
\begin{eqnarray} \label{eq:bounds_A}
 \lambda_{\min} \left( \left(\Aa - \Ab \right)^2 \right)  \leq \Delta(x^T,x)^2 \\
  \Delta(x^T,x) \leq \lambda_{\max} \left( \left(\Aa - \Ab \right)^2 \right) \nonumber
\end{eqnarray}
Finally, combining those bounds we can bound both solutions by 
{ \footnotesize
\begin{eqnarray} \label{eq:bounds_sol1_A} 
 \frac{1}{2} \left( \lambda_{\min} (\Aa + \Ab) - \sqrt{\lambda_{\max}\left( \left(\Aa - \Ab \right)^2 \right) + 4 p^2}  \right) \leq \lambda^- \\ 
 \leq \frac{1}{2} \left( \lambda_{\max} (\Aa + \Ab) -\sqrt{\lambda_{\max}\left( \left(\Aa - \Ab \right)^2 \right) + 4 p^2} \right). \nonumber
\end{eqnarray}
}
and
{ \footnotesize
\begin{eqnarray} \label{eq:bounds_sol2_A} 
 \frac{1}{2} \left( \lambda_{\min} (\Aa + \Ab) + \sqrt{\lambda_{\max}\left( \left(\Aa - \Ab \right)^2 \right) + 4 p^2} \right) \leq \lambda^+ \nonumber \\
 \leq \frac{1}{2} \left( \lambda_{\max}(\Aa + \Ab) + \sqrt{\lambda_{\max}\left( \left(\Aa - \Ab \right)^2 \right) + 4 p^2} \right), 
\end{eqnarray}
}
which asymptotically converge (as an approximation) to 
\begin{equation}
 p \pm \frac{\lambda_{\min} (\Aa + \Ab)}{2} \leq \lambda^\pm(x) \leq p \pm \frac{\lambda_{\max}(\Aa+\Ab)}{2}.
\end{equation}
In other words, the spectral density of the adjacency matrix is bimodal and part of the eigenvalues grows linearly with $p$, while the other part decreases at the same rate.

\subsubsection{Spectral properties as a function of the coupling $p$} \label{sec:A_funct_p}

\begin{figure}[!t]
\begin{center}
\includegraphics[width=0.98\linewidth]{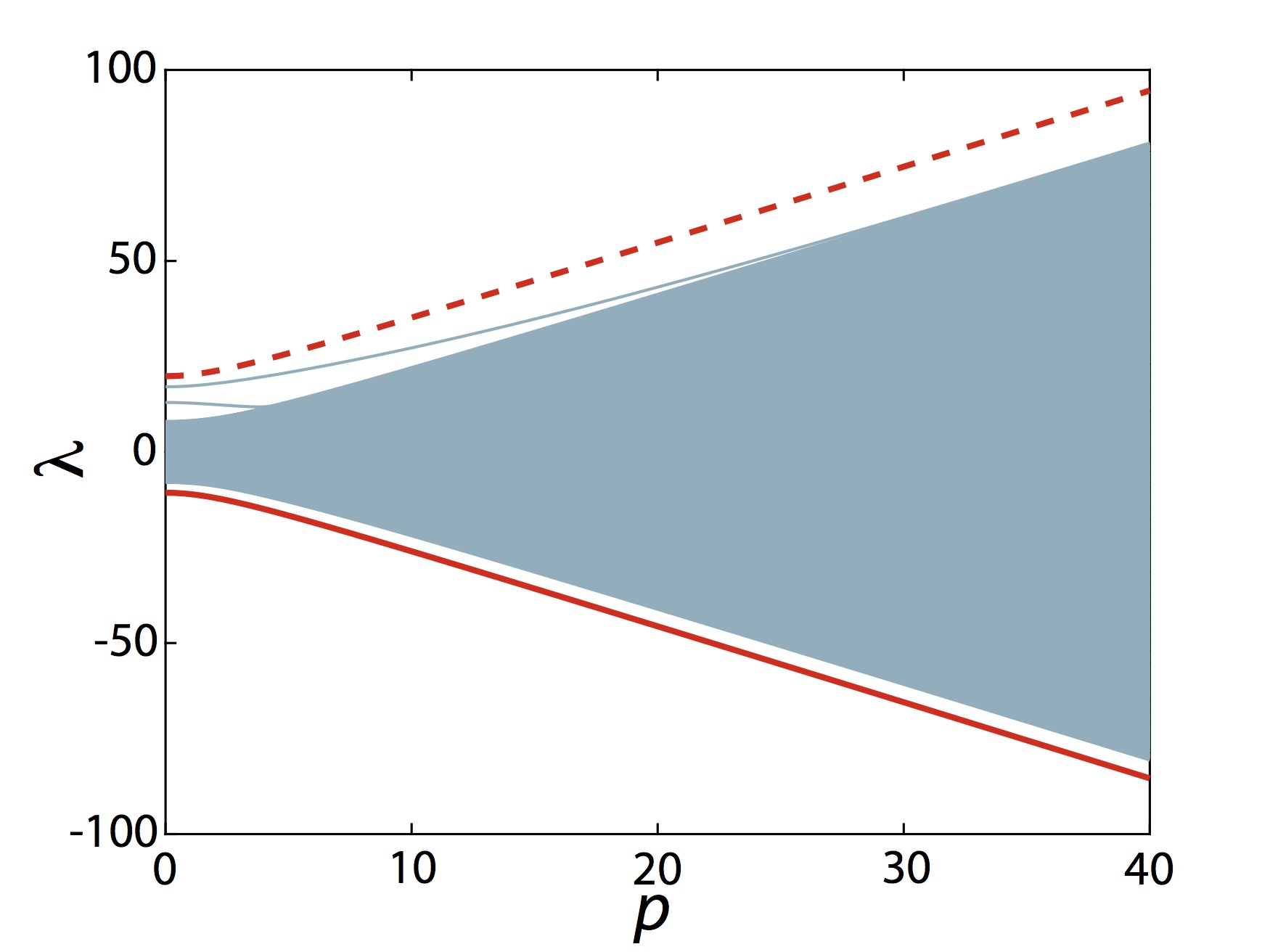}
\end{center}
\caption{Evaluation of the eigenvalues $\lambda(\Lap)$ as a function of the coupling parameter $p$ of a multiplex network composed by two Erd\"os -- Renyi layers with $n = 10^3$ nodes. The first layer has average degree $\E{k} = 12$, while the second has $\E{k} =16$. The coupling matrix is $\D = p \left( \frac{n}{\sum_i^n i}\text{diag}(1,2,...,n) \right)$. The dashed line is the adapted upper bound, while the continuous line is the adapted lower bound.}
\label{fig:BoundsAnon_identity}
\end{figure}

\begin{figure}[!t]
\begin{center}
\includegraphics[width=0.98\linewidth]{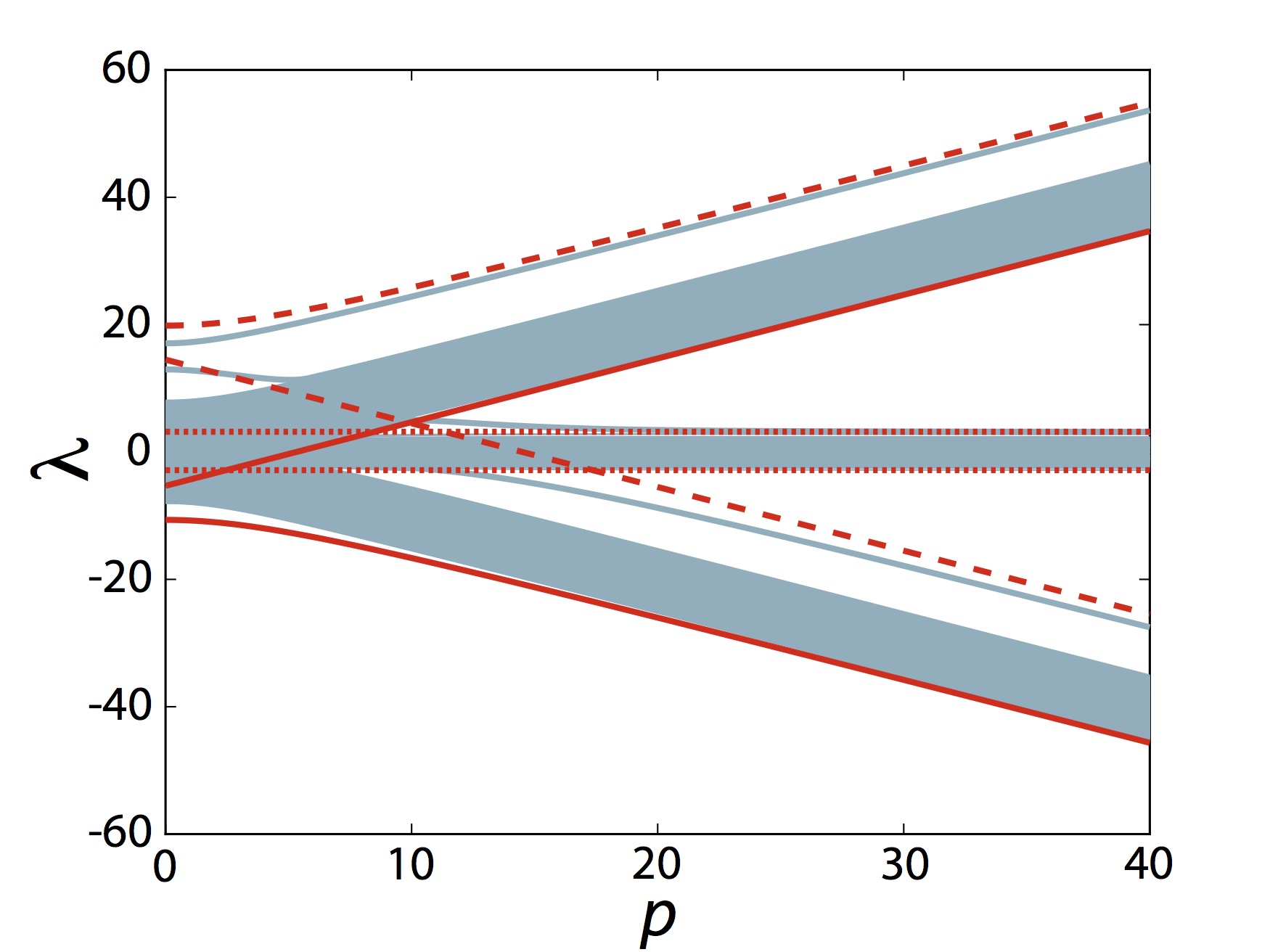}
\end{center}
\caption{Evaluation of the eigenvalues $\lambda(\A)$ as a function of the coupling parameter $p$ of a multiplex network composed by two Erd\"os -- Renyi layers with $n = 10^3$ nodes and a first layer with average degree $\E{k} = 12$ and a second has $\E{k} =16$. The coupling matrix is sparse. The dashed line is the adapted upper bound, the continuous line is the adapted lower bound and the dotted line was obtained numerically for the largest value of $p$ and shown as a reference.}
\label{fig:BoundsAsparse}
\end{figure}

In its general form, the first derivative is given as
\begin{equation} \label{eq:dldp_a}
  \dfrac{d \lambda}{dp} = \frac{2p y^T\D x}{\left( 2 \lambda y^T\D^{-1} x + b(y^T,x)\right)},
\end{equation}
where $x$ and $y^T$ are the right and left eigenvectors associated with $\lambda$. Firstly, focusing on the case where $\D = \I$ and using a similar approach as that applied to the Laplacian case, we can suppose that $\lambda$ is a constant function of $p$ or a function of $p^r$ with $r < 1$, however, it would give us $\dfrac{d \lambda}{dp} \sim p$, which is a contradiction. Next, we can suppose that it is a linear function of $p$, which implies $\dfrac{d \lambda}{dp} \rightarrow \pm 1$, depending on the sign of the linear coefficient. Finally, supposing that it is a function of $p^r$ with $r > 1$ we obtain that $\dfrac{d \lambda}{dp} \rightarrow 0$, since the denominator grows faster than the numerator, which again is a contradiction. In this way, based on such analysis we infer that the first derivative of $\lambda$ can assume only $\dfrac{d \lambda}{dp} \rightarrow \pm 1$. 

Secondly, for the general case observe that both, the numerator and the denominator of Equation \ref{eq:dldp_a} vary as a function of $\D$. Additionally, $\D$ weights the product of the components of the eigenvectors, which allows the derivatives to assume more values, even a ``continuous bulk'' instead of the bimodal distribution of the diagonal homogeneous case, similarly to the case discussed for the supra-Laplacian matrix. Here we also use the coupling matrix $\D = p \left( \frac{n}{\sum_i^n i}\text{diag}(1,2,...,n) \right)$. We show the spectral evolution as a function of $p$ for the non-homogeneous case in Figure~\ref{fig:BoundsAnon_identity}. Similarly to the Laplacian case, there are evidences that the bounds can be corrected using $\tilde{p} = \max\{ \text{diag} (\D)\} p$, hence
{ \footnotesize
\begin{eqnarray}
 \frac{1}{2} \left( \lambda_{\min} (\Aa + \Ab) - \sqrt{\lambda_{\max}\left( \left(\Aa - \Ab \right)^2 \right) + 4 \tilde{p}^2} \right) \leq \lambda \\
 \lambda \leq \frac{1}{2} \left( \lambda_{\max} (\Aa + \Ab) + \sqrt{\lambda_{\max}\left( \left(\Aa - \Ab \right)^2 \right) + 4 \tilde{p}^2} \right).
\end{eqnarray}}
Here we can also obtain a similar conclusion as for the supra-Laplacian case. From Figure \ref{fig:BoundsAnon_identity} we observe that the corrected bounds are not as close to the  homogeneous case.

Finally, we evaluate the sparse coupling case, whose analytical study was presented in Section \ref{sec:sparse}. As predicted, each uncoupled node implies a pair of eigenvalues that does not depend on $p$. In this way, if we have $\tilde{n}$ uncoupled nodes, the central part of the spectra will have $2\tilde{n}$ eigenvalues that do not change as a function of $p$. Next, $n - \tilde{n}$ grows linearly with $p$, while the other $n - \tilde{n}$ eigenvalues with $-p$. This is illustrated in Figure \ref{fig:BoundsAsparse}. Note that in this figure the horizontal lines bounding the central part of the spectra are not calculated, but numerically obtained and are only shown to serve as a reference.

\subsection{Probability transition matrix} \label{sec:app_P}

In this section, we evaluate the probability transition matrix, mainly focusing on its spectral properties as a function of the coupling parameter $p$. Due to the probabilistic nature of this matrix, we were not able to improve its bounds. Therefore, we mainly report numerical results.

Formally, the probability transition matrix is defined as
\begin{equation} \label{eq:prob_t}
\begin{split}
 \Pt = \hat{\D}^{-1} \A &= 
 \begin{bmatrix}
  (\tilde{\D}^A)^{-1} & \Om \\
  \Om & (\tilde{\D}^B)^{-1} 
 \end{bmatrix}
  \begin{bmatrix}
  \Aa & p\I \\
  p\I & \Ab 
 \end{bmatrix} = \\
 &= \begin{bmatrix}
  (\tilde{\D}^A)^{-1}\Aa & p(\tilde{\D}^A)^{-1} \\
  p(\tilde{\D}^B)^{-1} & (\tilde{\D}^B)^{-1} \Ab
 \end{bmatrix}
 \end{split}
\end{equation}
where $\tilde{\D}_{ii}^X = k_i^X + p$, and $X = \{ A, B\}$ represents the label of each layer. It is known that this matrix models the classical random walk, where the walker choses a neighbor based on the weights of its surrounding edges.

It is important to mention that in \cite{DeDomenico2014} the authors studied random walks on top of multiplex networks and analyzed them in terms of the normalized supra-Laplacian matrix. This matrix is defined as 
\begin{equation}
 \Lap^{\text{RW}} = \hat{\D}^{-1} \Lap = \I - \hat{\D}^{-1} \A = \I - \Pt,
\end{equation}
where $\Lap$ is the supra-Laplacian and $\Pt$ is the probability transition matrix. Note that the normalized supra-Laplacian matrix is intimately related to the probability transition matrix. In fact, their spectra are trivially related. Furthermore, we can also relate the spectra of the normalized Laplacian as follows
\begin{equation}
 \Lap^{\text{Norm}} =  \hat{\D}^{-\frac{1}{2}} \Lap \hat{\D}^{-\frac{1}{2}} = \I - \hat{\D}^{-\frac{1}{2}} \A \hat{\D}^{-\frac{1}{2}},
\end{equation}
where $\Sm = \hat{\D}^{-\frac{1}{2}} \A \hat{\D}^{-\frac{1}{2}}$ has the same set of eigenvalues as $\Pt$ and if $v$ is an eigenvector of $\Sm$, then $\hat{\D}^{-1} v$ is an eigenvector of $\Pt$ associated with the same eigenvalue \cite{Zhang2011}. Note, however, that $\Sm$ is symmetric \cite{Zhang2011}. In the context of random walks in multiplex networks, in \cite{Radicchi2014} the author used the normalized supra-Laplacian matrix. Here, in this section, we will study $\Pt$, defined in Equation \ref{eq:prob_t}.

Next, following our formalism, from Equation \ref{eq:prob_t}, we can define our QEP in its monic form as
{\scalefont{0.94}
\begin{eqnarray} \label{eq:p_mat_coef}
 \A &=& \I, \\
 \B &=& -\left( (\tilde{\D}^A)^{-1} \Aa + \Ab \tilde{\D}^B)^{-1} \right), \\
 \C &=& (\tilde{\D}^A)^{-1} \Aa \Ab (\tilde{\D}^B)^{-1} - (\tilde{\D}^A)^{-1} (\tilde{\D}^B)^{-1} p^2.
\end{eqnarray}
}
Note that such quadratic polynomial present some similarities with the one for the supra adjacency matrix, however the probability transition matrix is not symmetric and the matrices $\tilde{\D}^X$ presents a dependency on $p$. This fact, associated with the natural bound for stochastic matrices, make the derivation of the spectral bounds more complicated than the previous cases. Here we focus on the spectral properties of the probability transition matrix as a function of the coupling parameter $p$.

\subsubsection{Spectral properties as a function of the coupling $p$}

\begin{figure}[!t]
\begin{center}
\includegraphics[width=0.98\linewidth]{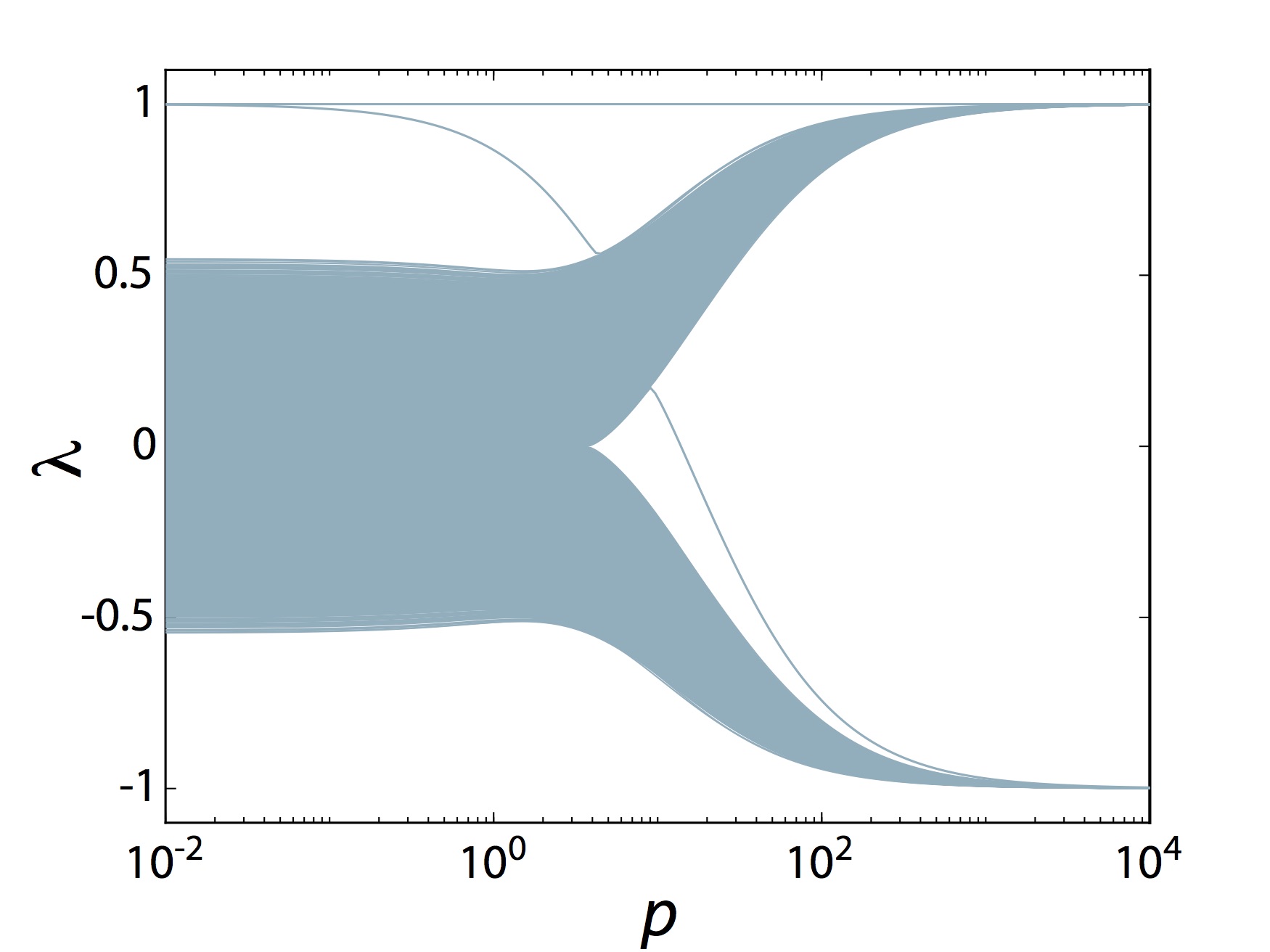}
\end{center}
\caption{Evaluation of the eigenvalues $\lambda(\Pt)$ as a function of the coupling parameter $p$ of a multiplex network composed by two Erd\"os -- Renyi layers with $n = 10^3$ nodes and a first layer with average degree $\E{k} = 12$ and a second with $\E{k} =16$.}
\label{fig:SpecPlog}
\end{figure}

For the sake of completeness, let us study the spectral properties of the transition matrix as a function of the coupling strength $p$.  This exercise is much more of an example than a practical application since we already know that the spectra are bounded on stochastic matrices, which does not allow unbounded grow. In Figure~\ref{fig:SpecPlog} we present the spectra as a function of the coupling parameter $p$. The first observation is that, aside from being bounded, the growth rate of the eigenvalues is quite different from what was observed for the Laplacian and adjacency cases. 

Firstly, lets proceed with the analysis of Equation \ref{eq:dldp} aiming for an approximation, which qualitatively describes the $\lambda(p)$. First of all, the partial derivative of $\B$ can be expressed as
\begin{equation} \label{eq:delB_pt}
 \frac{\partial \B}{\partial p} = -\frac{\partial}{\partial p} \left( \tilde{\D}^A \right)^{-1} \Aa - \frac{\partial }{\partial p} \left( \tilde{\D}^B \right)^{-1} \Ab,
\end{equation}
where the term
\begin{equation} \label{eq:delD_pt}
 \frac{\partial}{\partial p}  \left( \tilde{\D}^X \right)^{-1} = - \left( k_i^X + p \right)^{-2}.
\end{equation}
Next, expanding the partial derivative of $\C$ we have
\begin{eqnarray} \label{eq:delC_pt}
 \frac{\partial \C}{\partial p} = &+& \frac{\partial}{\partial p} \left( \tilde{\D}^A \right)^{-1} \Aa \Ab \left( \tilde{\D}^B \right)^{-1} + \nonumber \\
 &+& \left( \tilde{\D}^A \right)^{-1} \Aa \Ab \frac{\partial}{\partial p} \left( \tilde{\D}^B \right)^{-1} + \nonumber \\
 &-& \frac{\partial}{\partial p} \left( \tilde{\D}^A \right)^{-1} p^2 \left( \tilde{\D}^B \right)^{-1} + \nonumber \\
 &-& 2p\left( \tilde{\D}^A \right)^{-1} \left( \tilde{\D}^B \right)^{-1} +\nonumber \\
 &-& \left( \tilde{\D}^A \right)^{-1} p^2 \frac{\partial}{\partial p} \left( \tilde{\D}^B \right)^{-1}. 
\end{eqnarray}

All the expressions obtained so far are quite complicated to be analyzed in its exact form. Thus, we will proceed with an asymptotic analysis, aiming for a hypothesis of a possible formula that qualitatively describes the behavior of $\lambda_i$ as a function of $p$. In other words, we propose a formula that fits the expected asymptotic behavior, but we also expect it to work for smaller values of $p$. We must remark that this analysis is an approximation and, in order to verify its validity we perform numerical fittings and evaluate the obtained errors.

\begin{figure*}[!t]
\includegraphics[width=0.98\linewidth]{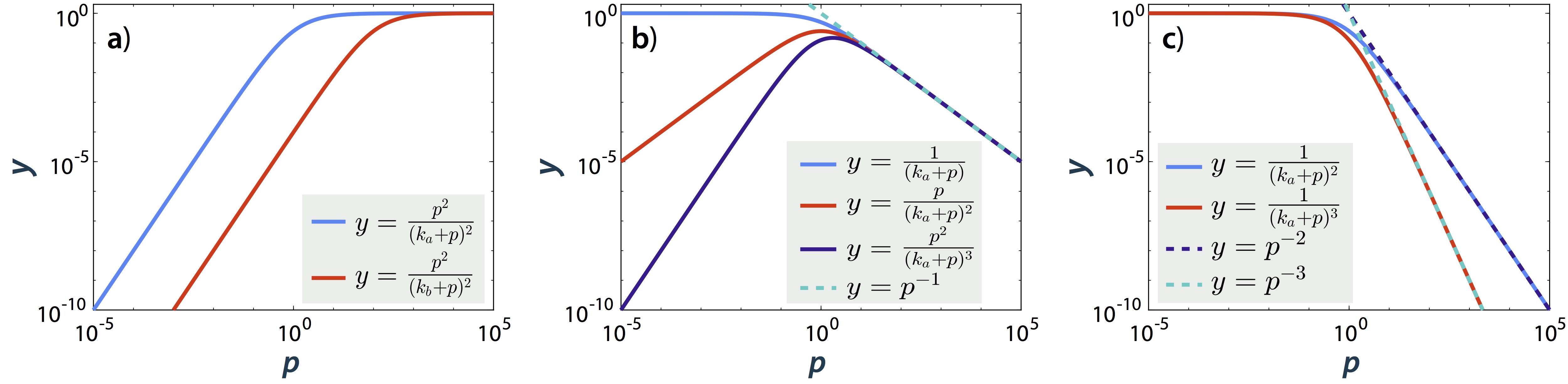}
\caption{Example of functions with different asymptotic behaviors. Functions in $O(1)$ in (a), $O(p^{-1})$ in (b) and in $O(p^{-2})$ and $O(p^{-3})$ in (c). In all plots we consider $k_a = 1$ and $k_b = 100$. Note that the asymptotic class does not change when we multiply the function by a constant.}
\label{fig:asymptotic_example}
\end{figure*}

From the previously mentioned perspectives, the asymptotic behavior are $\left( \frac{\partial \C}{\partial p} \right)_{ij} \in O(p^{-1})$, $\left( \frac{\partial \B}{\partial p} \right)_{ij} \in O(p^{-2})$, $\left( \B \right)_{ij} \in O(p^{-1})$ and, obviously, $\left( \A \right)_{ij} \in O(1)$. \footnote{Observe that, formally, if $f(x) = O(x^r)$, then $|f(x)| \geq K x^r$, where $K$ is a constant and $x \geq x_0$. Thus, multiplying $f(x)$ by a constant does not change its class. Besides, note that $O\left(x^{(r-1)} \right) \in O(x^r)$}. First of all, in Figure \ref{fig:asymptotic_example} we present some examples of functions with different asymptotic behaviors. In (a) we present a function in $O(1)$, showing that it can be bounded by a constant, in (b) we show some functions in $O(p^{-1})$, while in (c) two functions, one in $O(p^{-2})$ and the other in $O(p^{-3})$. Note that we can approximate some of the terms in equations \ref{eq:delB_pt}, \ref{eq:delD_pt} and \ref{eq:delC_pt} to the functions in Figure \ref{fig:asymptotic_example}. In Figure \ref{fig:asymptotic_example}, we also show examples of our guessed asymptotic behavior for
\begin{eqnarray*}
 &&\left( \tilde{\D}^A \right)^{-1} \sim \frac{1}{(k_x+p)} \in O(p^{-1}) \\
 &&2p\left( \tilde{\D}^A \right)^{-1} \left( \tilde{\D}^B \right)^{-1} \sim \frac{p}{(k_x+p)^2} \in O(p^{-1}) \\
 &&\left( \tilde{\D}^A \right)^{-1} p^2 \frac{\partial}{\partial p} \left( \tilde{\D}^B \right)^{-1} \sim \frac{p^2}{(k_x+p)^3} \in O(p^{-1})
\end{eqnarray*}
Next, in Figure \ref{fig:asymptotic_example} (c) we show examples for
\begin{eqnarray*}
 &&\frac{\partial}{\partial p}  \left( \tilde{\D}^X \right)^{-1} \sim \frac{1}{(k_x+p)^2} \in O(p^{-2}) \\
 &&\left( \tilde{\D}^A \right)^{-1} \Aa \Ab \frac{\partial}{\partial p} \left( \tilde{\D}^B \right)^{-1} \sim \frac{1}{(k_x+p)^3} \in O(p^{-3}).
\end{eqnarray*}
Note that we considered a single value of $k_x$, without considering products between different constants. Besides, just one term is considered. We remark that the main goal of this exercise is to have insights on the qualitative behavior of more complicated functions, such as Equation \ref{eq:delC_pt}. In other words, the performed approximations are not expected to quantitatively predict those terms, but qualitative represent and ``catch'' the main behavior of those functions.

Firstly, recall that the spectra on stochastic matrices is bounded, which consequently restricts its derivatives. In other words, $\lambda_i \in O(1)$. However, for the sake of the argument, let us suppose that $\lambda_i = c_1 p^r + O(p^{r-1})$, hence $\dfrac{d \lambda_i}{dp} = c_1 r p^{(r-1)} + O(p^{r-2})$, where $r$ is an integer. Thus, comparing with Equation \ref{eq:dldp}, we have 
\begin{eqnarray} \label{eq:dldp_Pt_asy}
 \dfrac{d \lambda_i}{dp} &=& c_1 r p^{(r-1)} + O(p^{r-2}) =  \\
 &=& \frac{\left( - 2c_1 p^r + O(p^{r-1}) \right) \times O(p^{-2}) + O(p^{-1})}{ c_1 p^r + O(p^{r-1}) + O(p^{-1})}, \nonumber
\end{eqnarray}
that can be rewritten as
\begin{eqnarray}
 &&\left( c_1 r p^{(r-1)}+ O(p^{r-2}) \right) \left( c_1 p^r + O(p^{r-1}) + O(p^{-1}) \right) =  \nonumber \\
 &&= \left( - c_1 p^r + O(p^{r-1}) \right) \times O(p^{-2}) + O(p^{-1}),
\end{eqnarray}
which simplifies to
\begin{eqnarray}
 &&c_1^2 r p^{(2r-1)} + O(p^{2r-2}) + O(p^{-1}) = \nonumber \\
 &&= c_2 p^{r-2} + O(p^{r-3}) + O(p^{-1}),
\end{eqnarray}
which implies that $\dfrac{d \lambda}{dp} \rightarrow 0$ and $r \leq 0$ since on the left-hand side we have a function in $O(\max \{2r-1, -1 \})$, while, on the right-hand side we have a function in $O(\max \{r-2, -1 \})$. This simple analysis suggests that $r \leq 0$, for consistency. Note that we are not inferring anything regarding its ``velocity'' (how fast it goes to zero). Such arguments reinforce that $\lambda_i \in O(1)$, as previously mentioned. However, there are a huge class of functions that satisfies such restriction. In order to satisfy the so far established restrictions, let us suppose that 
\begin{equation}  \label{eq:lambda_pt_val}
 \hat{\lambda_i} = \frac{k_0 p^2}{(p+c_0)^2} + \sum_{k=1}^K \frac{\tilde{c}_k p^{k-1}}{(p + c_k)^k} = \frac{k_0 p^2}{(p+c_0)^2} + O(p^{-1}),
\end{equation}
which is a function that satisfies our previous analysis. Thus, it also implies that
\begin{eqnarray} \label{eq:d_lambda_pt_val}
 \dfrac{d \hat{\lambda_i}}{dp} &=& \frac{2 k_0 c_0 p}{(p+c_0)^3} - \sum_{k=1}^K \frac{\tilde{c}_k p^{k-2}( c_k - k c_k + p)}{(p + c_k)^{k+1}} = \\
 &=& \frac{2 k_0 c_0 p}{(p+c_0)^3} - \sum_{k=1}^K \frac{\tilde{c}_k p^{k-1}}{(p + c_k)^{k+1}}  + O(p^{-3}),
\end{eqnarray}
which yields $\dfrac{d \hat{\lambda_i}}{dp} \in O(p^{-2})$. Next, from Equation~\ref{eq:dldp} we have
\begin{eqnarray} \label{eq:dldp_Pt_asy2}
 \dfrac{d \lambda_i}{dp} \sim \frac{O(1) \times O(p^{-2}) + O(p^{-1})}{O(1) \times O(1) + O(p^{-1})} = \frac{O(p^{-1})}{O(1)} \in O(p^{-1}).
\end{eqnarray}
Note that it allows a set of possible solutions and, among them, it allows our initial supposition, Equation \ref{eq:lambda_pt_val}, i.e., $\lambda_i \sim \hat{\lambda_i}$. Note that $\hat{\lambda_i} \in O(1)$ and $\dfrac{d \hat{\lambda_i}}{dp} \in O(p^{-2})$, which is also in $O(p^{-1})$, as expected from Equation \ref{eq:dldp_Pt_asy2}. Besides, for the sake of visualization, in Figure \ref{fig:asymptotic_example} (a) we show two examples of the leading term of Equation \ref{eq:lambda_pt_val}.

\begin{figure}[!t]
\begin{center}
\includegraphics[width=0.98\linewidth]{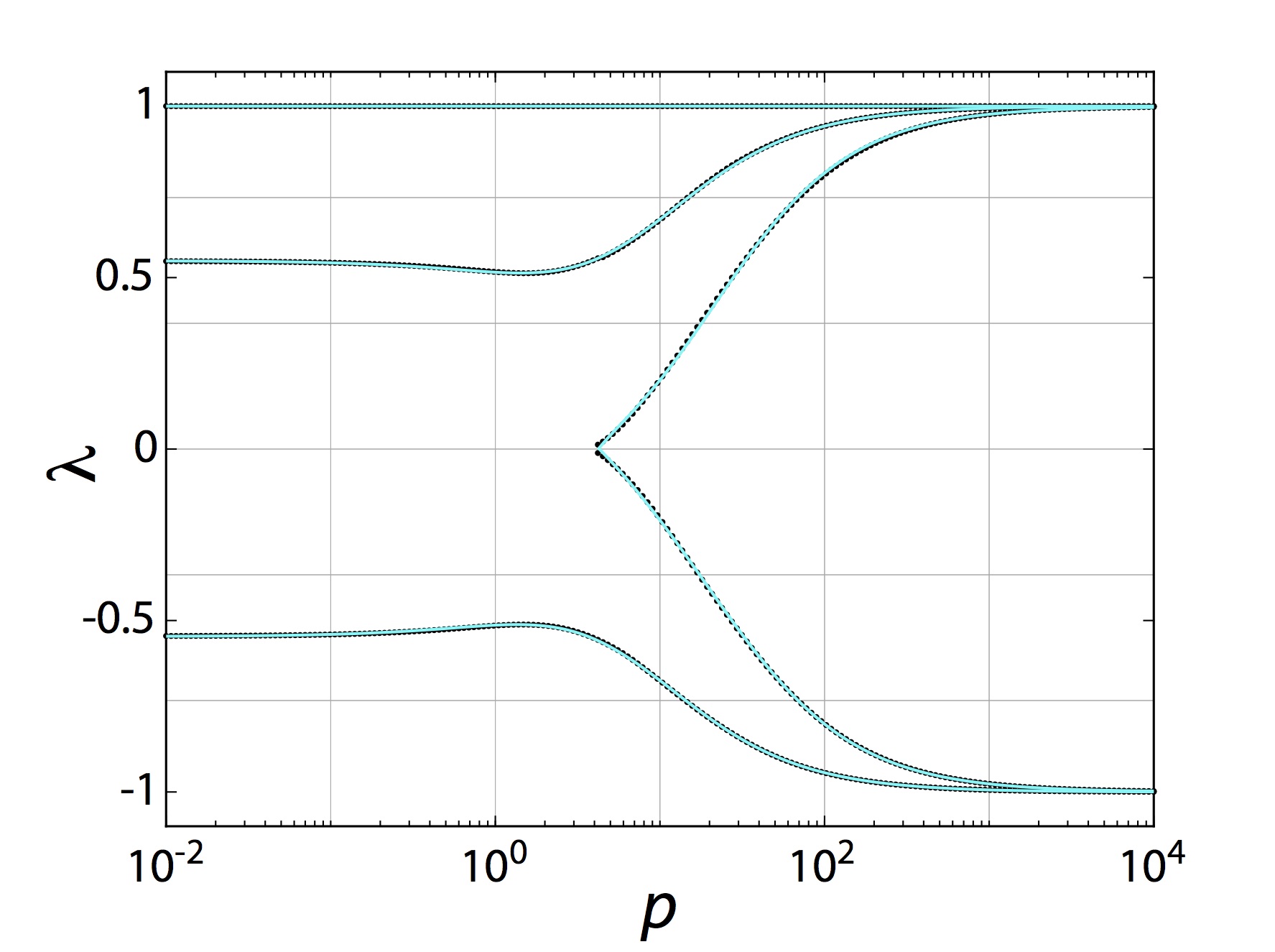}
\end{center}
\caption{Evaluation of the eigenvalues of the probability transition matrix, $\lambda_i(p)$, for $i=N$ and $i=3$ (or $i=2$ after the crossing, in order to have a continuous curve) as a function of the coupling parameter $p$ of a multiplex network composed by two Erd\"os -- Renyi layers with $n = 10^3$ nodes and a first layer with average degree $\E{k} = 12$ and a second with $\E{k} =16$. The dots are the obtained eigenvalues from eigendecomposition of $\Pt$, while the continuous red lines are the fitted curves from Equation~\ref{eq:lambda_pt_val}, where we used just the first term on the summation, i.e., $K=1$.}
\label{fig:SpecPlogFit}
\end{figure}

\begin{table*}[ht!]
\caption{Parameter values of the network reported in Figure~\ref{fig:SpecPlogFit}. The confidence intervals are given in parenthesis and the goodness of fit is measured by the Sum of Squares Due to Error (SSE).}
\label{Table1} 
\begin{tabular}{|l|l|l|l|l|l|}
\hline
Eigenvalue & $k_0$ & $c_0$ & $c_k$ & $\tilde{c}_k$ & SSE\\
\hline
$\lambda_{1} = 1$             & $1  (1, 1)$   & $< 10^{-5}$ & $< 10^{-5}$ & $< 10^{-5}$ & $< 10^{-6}$ \\
$\lambda_{3}$             & $0.9999  (0.9999, 1)$   & $5.514  (5.502, 5.526)$ & $8.618  (8.579, 8.657)$ & $4.732  (4.712, 4.753)$ &  $< 10^{-4}$ \\
$\lambda_{n-1}$             & $0.9985  (0.9975, 0.9996)$   & $11.27  (11.18, 11.36)$ & $-2.141  (-2.629, -1.653)$ &  $-0.1541  (-0.1839, -0.1242)$ & $< 10^{-2}$ \\
$\lambda_{n+1}$             & $-0.9997  (-0.9999, -0.9995)$   & $0.1273  (0.1186, 0.1361)$ & $20.73  (20.61, 20.85)$ &  $23.58  (23.53, 23.63)$ & $<10^{-4}$ \\
$\lambda_{N}$             & $-1  (-1, -1)$   & $5.344  (5.334, 5.353)$ & $8.444  (8.412, 8.477)$ &  $-4.615  (-4.632, -4.598)$ & $< 10^{-4}$ \\
\hline
\end{tabular}
\end{table*}

\begin{figure}[!t]
\begin{center}
\includegraphics[width=0.98\linewidth]{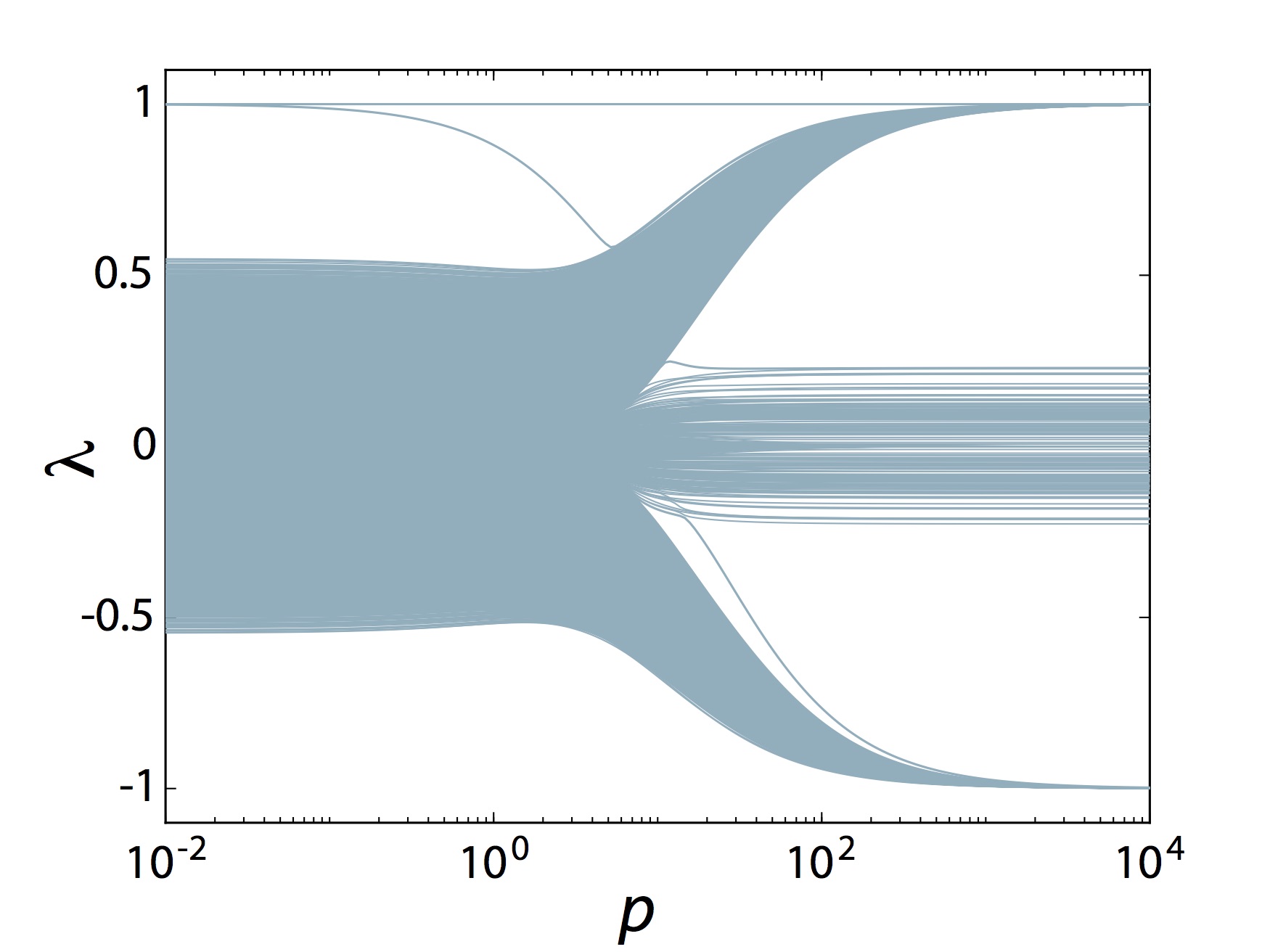}
\end{center}
\caption{Evaluation of the eigenvalues $\lambda(\Pt)$ as a function of the coupling parameter $p$ of a multiplex network composed by two Erd\"os -- Renyi layers with $n = 10^3$ nodes, a first layer with average degree $\E{k} = 12$ and a second with $\E{k} =16$. The coupling matrix is sparse.}
\label{fig:SpecPlogSparse}
\end{figure}

\begin{figure}[!t]
\begin{center}
\includegraphics[width=0.98\linewidth]{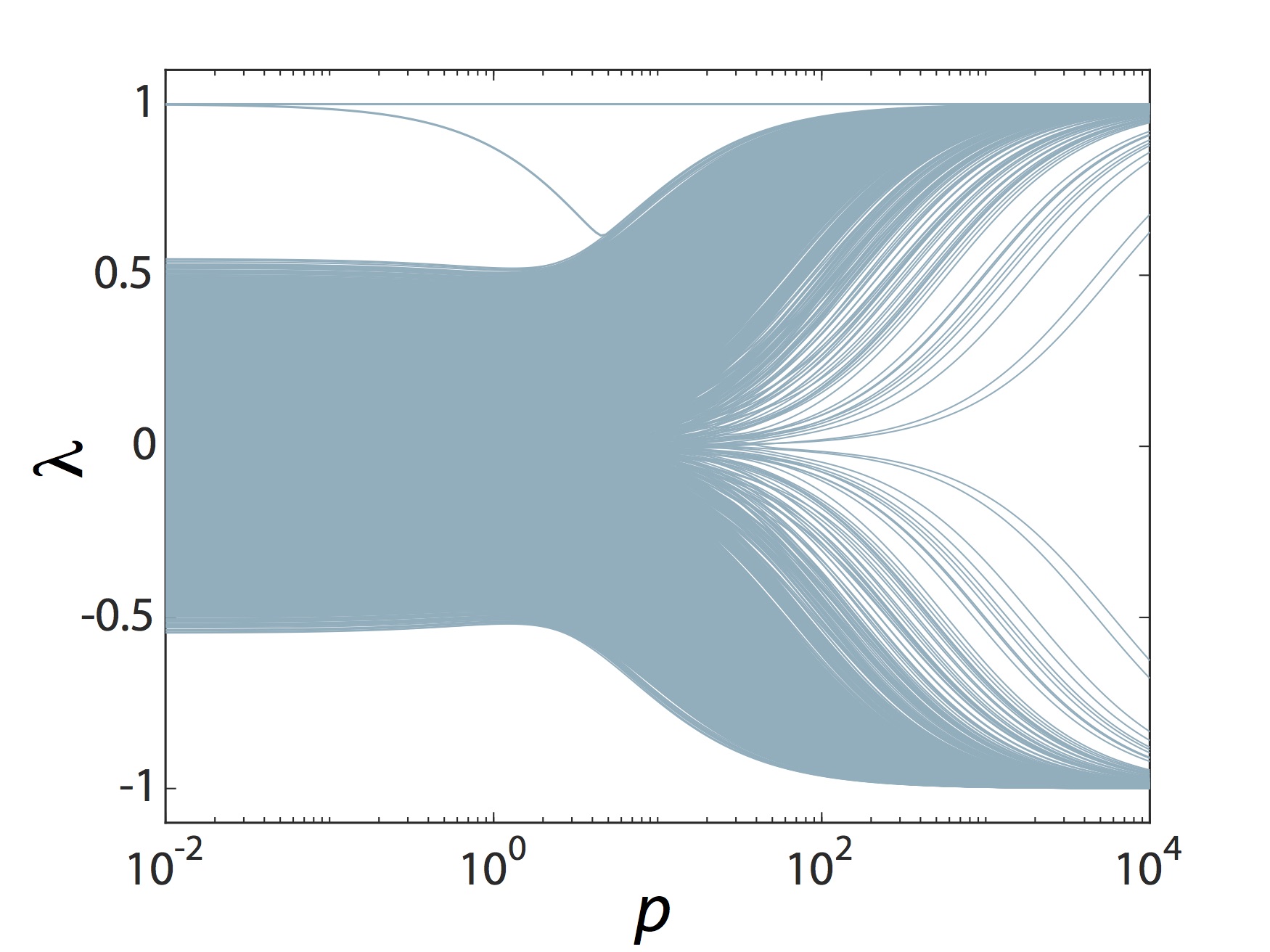}
\end{center}
\caption{Evaluation of the eigenvalues $\lambda(\Pt)$ as a function of the coupling parameter $p$ of a multiplex network composed by two Erd\"os -- Renyi layers with $n = 10^3$ nodes. The first layer with average degree $\E{k} = 12$ and the second has $\E{k} =16$. The coupling matrix is $\D = p \left( \frac{n}{\sum_i^n i}\text{diag}(1,2,...,n) \right)$.}
\label{fig:SpecPlogHetero}
\end{figure}

Next, we proceed with a numerical experiment, extracting some eigenvalues presented in Figure~\ref{fig:SpecPlog} we perform a fitting aiming to obtain the same curve. We chose 5 eigenvalues: (i) the leading eigenvalue, $\lambda_{1} = 1$, just as a reference and to emphasize that our proposed equation also works for that case, (ii) $\lambda_{3}$, the first eigenvalue on the bulk (note that there can be a crossing between $\lambda_{3}$ and  $\lambda_{2}$, which would change the index of the eigenvalue -- here we are not going to enter into details of this possible crossing behavior and, in order to avoid that, we chose to follow the third eigenvalue), (iii) $\lambda_{N}$, the smallest eigenvalue and (iv) the two intermediate eigenvalues $\lambda_{n-1}$ and $\lambda_{n+1}$, where we just considered their values after the spectra divides into two parts. Note that the error of these two curves is expected to be larger than the previous cases. It is important to remark that, as previously mentioned, there can be a crossing between $\lambda_{3}$ and  $\lambda_{2}$, but here we are looking at the main global behavior and such a change would not be a big source of error. In this way, we are showing that there is a set of parameters that approximates the spectra using the proposed equations. The proposed mentioned experiment does not serve as a proof, but it does serve as an evidence of such, or a similar, behavior. 

Following the proposed pipeline, firstly, in order to obtain the fittings, we used the nonlinear least squares method, the Levenberg-Marquardt algorithm \cite{Levenberg1944,Marquardt1963,More1978} and the least absolute residual (LAR) robust regression. Additionally, all the initial conditions were set to one. In Figure~\ref{fig:SpecPlogFit} we show the obtained fittings and the numerically obtained eigenvalues. Complementary, in Table~\ref{Table1} we present the fitted parameters. Interestingly, we observe that the proposed approximation fits really well the observed curves, which can be objectively measured by means of the Sum of Squares Due to Error (SSE), whose values are also reported in Table~\ref{Table1}. Thus, the behavior of $\lambda_i$ assumed in Equation~\ref{eq:lambda_pt_val} seems to be a very good guess. Besides, we also observe that there seems to be a symmetry on the obtained parameters for $\lambda_{3}$ and $\lambda_{N}$, which are close. The only exception is $k_0$, since both have the same modulus, but with a different sign, as expected. The last important observation also regards the parameter $|k_0|$. Note that such a parameter is very close to one on all the fittings, suggesting some underlying property of our formulation.

Finally, for the sake of completeness and for comparison reasons, we numerically evaluate the spectra of the probability transition matrix for the sparse and heterogeneous coupling cases. Regarding the sparsity, in Figure~\ref{fig:SpecPlogSparse}, we present a similar experiment as done for the supra-Laplacian and supra-adjacency cases. Similarly to those experiments, here we also observe a group of eigenvalues that do not change as a function of $p$. Moreover, we also verified that for $\hat{n} = 100$ decoupled nodes, we have $2\hat{n} = 200$ eigenvalues that remain constant, validating the insights we obtained in Section~\ref{sec:sparse}. Although the behavior observed is similar to the previously studied matrices, for the probability transition matrix we observe a slightly different behavior for intermediate values of $p$ (here $1<p<10$), where the intermediate eigenvalues change, forming the ``central bulk''. 

Furthermore, we remark that for the heterogeneous coupling ($\D = p \left( \frac{n}{\sum_i^n i}\text{diag}(1,2,...,n) \right)$), if compared with the supra-Laplacian and supra-adjacency, a completely different behavior emerged. In the probability transition matrix case, the spectra seem to be always bi-modal. This effect is shown in Figure \ref{fig:SpecPlogHetero}, where, for a large enough value of $p$, the eigenvalues tend to a constant. It is also noteworthy that the rate at which this phenomenon takes place is much slower than the rate of the homogeneous case, shown in Figure \ref{fig:SpecPlog}.

\section{Discussion and conclusions} \label{sec:discussion} 

From the developed theory, we applied and analyzed three different matrices: (i) the supra-Laplacian, (ii) the supra-adjacency and (iii) the probability transition matrix. In all these cases we have considered three different coupling schemes: (a) diagonal homogeneous coupling, $\D = p\I$, (b) diagonal homogeneous sparse coupling and (c) diagonal heterogeneous coupling, $\D = p \left( \frac{n}{\sum_i^n i}\text{diag}(1,2,...,n) \right)$. Regarding the supra-Laplacian and the supra-adjacency matrices, on the first scenario, (a), we were able to extract some analytical results regarding the derivatives of the eigenvalues, which suggested a different behavior for the other two cases, (b) and (c). On the other hand, regarding the probability transition matrix, due to its stochastic nature, we were not able to go further with the analytical analysis. However, we followed an asymptotic analysis, proposing a function that describes the eigenvalues behavior. This function was validated with numerical fittings of the original spectra. Although it is just an approximation, it also helps us understand the nature of the phenomena behind this structure. Furthermore, we also reported the differences between the spectral distributions for large $p$, where we can have bimodal, multi-modal or even a continuous bulk for the adjacency and Laplacian cases, just changing the coupling matrices. On the sparse case, this analysis was analytically supported, while the other cases were explored numerically.

Our analysis pointed out some important features about multiplex systems. As a general observation, as we increase $p$ we will find (roughly) three different structural phases, which might take place at different points for each structure and matrix.  Thus, the structural phases of a multiplex network can be defined as: (i) decoupled phase, for small values of $p$, where the layers are virtually decoupled and act by themselves, with a neglectable interaction, (ii) multiplex/multilayer phase, where the system is coupled and the intra-layer edges play an important role and (iii) a network of layers phase, where the structure of the network of layers plays the major role. Note that, from the perturbation theory point of view, in the decoupled phase the eigenvalues are basically the union of the eigenvalues of the individual layers plus some perturbations. On the other extreme, in the network of layers phase, the intra-layer edges might be understood as the perturbation, since $p \gg 1$. In this case, we can interpret the system as a set of $n$ virtually disconnected small networks, whose structure is given by the network of layers (considering a multiplex case, where each node has a counterpart on the other layers). Finally, the most interesting scenario is the multiplex phase, where the inter and intra-layer topologies play a fundamental role on the dynamics.

Throughout our analysis, we were able to verify these regimes in the different matrices we evaluated. Although all of them showed this behavior, the differences between those matrices are also evident. Considering the supra-Laplacian and supra-adjacency matrices with diagonal homogeneous coupling, we observe that, for a large enough $p$, the spectral distribution is bi-modal, while for the sparse case we have three bulk's, where the central one results from the nodes that do not have an inter-layer edge. Finally, on the diagonal heterogeneous coupling, we observe a completely different behavior, where the eigenvalues are distributed into a single bulk.  We remark that, although we were not able to analytically quantify this last phenomenon, our analysis suggested such a behavior. 

Furthermore, comparing those results with the ones obtained using the probability transition matrix, we observed a completely different behavior. On the diagonal, homogeneous or heterogeneous cases, the spectra seem to be bi-modal for a sufficiently large $p$. Note that for the homogeneous case this convergence to the bulks is much faster than the heterogeneous case. This is an interesting phenomenon since it contrasts with the supra-adjacency and supra-Laplacian cases, where the heterogeneous coupling implies a ``continuous'' bulk. It is noteworthy that our predictions for a central bulk for the uncoupled nodes are also fulfilled for the probability transition matrix.

The analysis performed here emphasize the importance of a proper study of the structural phases in different contexts.  The sparse and heterogeneous cases might change completely the spectra (depending on $p$). Obviously, the analysis should also take into account the correct matrix since the structural changes are different from case to case. In other words, different matrices present their phases in different intervals (values of $p$). In this context, dynamical insights can also be useful for a better understanding.

Since all the analyzed matrices are also related to dynamical processes, the results reported here will directly impact on these processes too. Note that in \cite{GomezPRL2013} the authors analyzed diffusion processes in multiplex networks and found the so-called superdiffusion. This process is described by the supra-Laplacian matrix and it is intrinsically connected to the so-called structural transition of this matrix as pointed in \cite{GomezPRL2013} and latter discussed in \cite{Mieghem2015}, where the authors found the exact structural transition point. Furthermore, in \cite{Arruda2017}, while studying epidemic spreading in multiplex networks, the authors verified this structural behavior in the analysis of the supra-adjacency matrix. Besides, it was also shown that it is intimately related to spreading processes and the layer-localization phenomena \cite{Arruda2017}. Thus, in the mentioned cases, the dynamical regimes can be understood as a consequence of the structural changes.

In summary, we have proposed a new mathematical formalism for the analysis of spectral properties in multiplex networks using the polynomial eigenvalue problem. This approach, we reduces the dimensionality of our matrices (coefficient matrices) at the cost of a higher order of the characteristic polynomial. This technique might seem counterintuitive at a first glance, but it reveals an underlying relationship between the eigenvalues of the matrices associated to multiplex structures. In contrast to single layer networks, multiplex networks are defined as matrix functions since we are interested in weighting inter and intra-layer edges differently. Thus, they depend on the coupling parameter, here denoted by $p$. Therefore, it is of utmost importance to derive spectral bounds as a function of $p$ and to obtain insights on their asymptotic behavior.

We hope this work motivates the community to study in further details the structural behavior of multiplex and multilayer systems and their dynamical consequences. Besides, other matrices and processes might also be studied and evaluated. In this case, we believe that our formalism might also be helpful. Finally, we also hope to motivate studies on the analysis of eigenvectors, which is still lacking in the literature and are of great importance as they were shown to play a major role in dynamical processes \cite{Goltsev2012,Arruda2017}.

\textbf{Acknowledgment}
FAR acknowledge CNPq (grant 307974/2013-8) and Fapesp (grant 2013/26416-9) for financial support. GFA acknowledges Fapesp for the sponsorship provided (grants 2012/25219-2). YM acknowledges partial support from the Government of Aragón, Spain through grant E36-17R, and by MINECO and FEDER funds (grant FIS2017-87519-P).

\appendix
\section{Jordan triple} \label{sec:app_jordan}

The definitions presented in this section are not studied in details here since our goal is focused on applications, however, we refer the reader to \cite{Lancaster:book, gohberg1982matrix} for more information on this class of problems. Here we reproduce some important definitions that might me useful for some readers, allowing them to extend the results presented in this work to other contexts.

\theoremstyle{definition}
\begin{definition}{Jordan triple \cite{Lancaster:book}}:
 Denoting by $\J$ the Jordan matrix, $(\X,\J,\Y)$ is the Jordan triple of $\Q(\lambda)$, where $\X$ and $\Y$ are the right and left eigenvectors. We also have that $(\Y^*,\J^*,\X^*)$ is the Jordan triple of $\Q(\lambda)^*$. Note that the Jordan matrix, $\J$, the diagonal blocks are the Jordan blocks. Besides, if all the eigenvalues are simple, $\J$ is a diagonal matrix, where $\J_{ii} = \lambda_i$. Additionally, observe that $\X \in \mathbb{R}^{n \times nl}$, where $\X = \left[ x_1, ..., x_{nl}\right]$ is composed by the right eigenvectors, while $\Y \in \mathbb{R}^{nl \times n}$, where $\Y = \left[ y_1, ..., y_{nl}\right]^T$ is composed by its left eigenvectors. Finally, $\J \in \mathbb{R}^{nl \times nl}$.
\end{definition}

\theoremstyle{definition}
\begin{definition}{The left set of eigenvectors $\Y$ \cite{Lancaster:book}}:
 The left eigenvectors can be defined in terms of the right eigenvectors and the matrix $\J$ as
 \begin{equation}
  \Y = \begin{bmatrix}
        \X \\
        \X \J \\
    \vdots  \\
        \X \J^{l-1} 
       \end{bmatrix}^{-1}
       \begin{bmatrix}
        \Om \\
        \vdots  \\
        \Om \\
        \I
       \end{bmatrix}
 \end{equation}
\end{definition}

\theoremstyle{definition}
\begin{definition}{Useful relations for the order two polynomial eigenvalue problem}:
\begin{eqnarray}
  &\X\Y = \Om, \\
  &\X\J\Y = \I \\
  &\X \J^2 + \B \X \J + \C \X = \Om \\
  &\J^2 \Y + \J \Y \B + \Y \C = \Om
\end{eqnarray}
\end{definition}

\bibliographystyle{apsrev}
\bibliography{references}

\end{document}